\documentclass[twocolumn,superscriptaddress]{revtex4-1}
\usepackage{amsmath}
\usepackage{graphicx}
\usepackage{xcolor}
\usepackage{braket}
\usepackage{tikz}
\usepackage{tikz-3dplot}
\DeclareMathOperator*{\argmin}{arg\,min} 

\newcommand{\bR}{\textbf{R}}

\begin{document} 

\title{Excited states in variational Monte Carlo using a penalty method}

\author{Shivesh Pathak}
\affiliation{University of Illinois at Urbana-Champaign, Urbana, IL 61801} 
\author{Brian Busemeyer}
\affiliation{Center for Computational Quantum Physics, Flatiron Institute, York, NY 10010}
\author{Jo\~ao N. B. Rodrigues}
\affiliation{Universidade Federal do ABC, Santo Andr\'e, SP, 09210-580, Brazil}
\author{Lucas K. Wagner}
\email{lkwagner@illinois.edu}
\affiliation{University of Illinois at Urbana-Champaign, Urbana, IL 61801}

\begin{abstract}
The authors present a technique using variational Monte Carlo to solve for excited states of electronic systems. 
The technique is based on enforcing orthogonality to lower energy states, which results in a simple variational principle for the excited states.
Energy optimization is then used to solve for the excited states.
This technique is applied to the well-characterized benzene molecule, in which $\sim$10,000 parameters are optimized for the first 12 excited states.
Agreement within approximately 0.2 eV is obtained with higher scaling coupled cluster methods; small disagreements with experiment are likely due to vibrational effects.
\end{abstract}

\maketitle

\section{Introduction}

The ground and first few excited states determine the behavior of most systems in materials, condensed matter and chemistry. 
A hallmark of correlated electron physics is the presence of many disparate excited states near the ground states, which may differ in complex ways that depend on exactly how electronic correlation is treated, as has been shown in the Hubbard model\cite{simonscollaborationonthemany-electronproblemSolutionsTwoDimensionalHubbard2015}.
Methods to access these low energy states in strongly correlated systems are fundamental to understanding them. 
There is thus a need for scalable methods that can treat strongly correlated systems and non-perturbatively access excited states, which would generate new insights into long-standing challenging systems such as the low-energy behavior of high-Tc cuprates/pnictides and twisted bilayer graphene. 

Quantum Monte Carlo techniques such as variational and diffusion Monte Carlo\cite{foulkesQuantumMonteCarlo2001} are based on many-body, non-perturbative wave functions. 
They offer low scaling, typically of order ${\cal O}(N_e^{3-4})$, where $N_e$ is the number of electrons, and can offer impressive accuracy for that low scaling.\cite{simonscollaborationonthemany-electronproblemDirectComparisonManyBody2020}
However, these techniques are most developed for ground state calculations. 
By far the most common technique to approximate excited states is to fix a single Slater determinant or linear combination of Slater determinants to approximate the ground state, while using a Jastrow\cite{jastrowManyBodyProblemStrong1955} factor and diffusion Monte Carlo to address some of the electron correlation.
This is the technique outlined in Ref~\cite{foulkesQuantumMonteCarlo2001}, and similar techniques are used in auxilliary field quantum Monte Carlo.\cite{maExcitedStateCalculations2013}
This technique works surprisingly well for many materials.\cite{mitasQuantumMonteCarlo1994a,williamsonDiffusionQuantumMonte1998,schillerPhaseStabilityProperties2015, zhengComputationCorrelatedMetalInsulator2015, wagnerDiscoveringCorrelatedFermions2016,frankManyBodyQuantumMonte2019, wangBindingExcitationsSi2020}
However, this technique depends on the quality of the wave function that generated the fixed trial function; if correlation included by the Jastrow would change the optimal Slater part of the wave function, then the \textit{ansatz} is suboptimal, as demonstrated clearly in Ref~\cite{tranImprovingExcitedState2020}. 

Recently, and not so recently, there have been extensions to the variational and diffusion Monte Carlo (VMC, DMC) methods to address excited states systematically.
These techniques seek to improve the accuracy of excited states over the simple method explained above by optimizing the antisymmetric part of the wave function. 
Bernu and Ceperley proposed an algorithm based on diffusion Monte Carlo,\cite{ceperleyCalculationExcitedState1988} which has not been applied to many practical cases, since it has a sign problem which leads to exponential scaling in the system size.
Blunt et al.\cite{bluntExcitedstateApproachFull2015} proposed an algorithm using full CI configuration interaction, which is also exponentially scaling.
On the low-scaling side, Filippi and coworkers\cite{doi:10.1021/ct900227j, doi:10.1021/acs.jctc.9b00476, doi:10.1021/ct1006295} implemented a method similar to state averaged CASSCF in VMC, and demonstrated the technique on impressively large wave function expansions. 
However, state averaging is not optimal when the optimal ground and excited orbitals are very different,\cite{pinedafloresExcitedStateSpecific2019} as often occurs in strongly correlated systems. 
Neuscamman and coworkers\cite{doi:10.1021/acs.jctc.6b00508, doi:10.1021/acs.jctc.7b00923, doi:10.1021/acs.jpca.8b10671, doi:10.1021/acs.jctc.8b00879} have proposed a low-scaling method that instead uses alternate objective functions to optimize excited states. 
While this technique does not suffer from the state averaging problem, it so far has only been applied to very few excited states, and can experience difficulties converging to the correct state.\cite{cuzzocreaVariationalPrinciplesQuantum2020}
Finally, Choo et al.\cite{chooSymmetriesManyBodyExcitations2018} have used an orthogonalization step after each optimization step to access excited states on a lattice, but the algorithm has not been demonstrated on first principles models.

In this manuscript, we implement and demonstrate a simple penalty method based on orthogonalizing to lower energy states to compute excited states using variational Monte Carlo.
Similar techniques are commonly used in density matrix renormalization group calculations,\cite{stoudenmireStudyingTwoDimensionalSystems2012} but to our knowledge have not been applied in the variational Monte Carlo context. 
This technique obtains excited states one by one by enforcing orthogonality to lower energy states, and can optimize general wave function parameters, including orbital parameters as shown here. 
The scaling of the technique is $\mathcal{O}(N_{ex} N_e^M) + c \mathcal{O}(N_{ex}^2 N_e^M)$, where $N_e$ is the number of electrons, $N_{ex}$ is the number of excited states computed, $c$ is a small constant, and $M$ is dependent on the wave function, 3 for a Slater-Jastrow wave function. 
The orthogonalization based technique allows for access to multiple excited states, and does not require state averaging.
The technique is implemented in the \texttt{pyqmc} package, available online.\cite{pyqmc}
The method is applied to benzene with a $\sim$10,000 parameter wave function, showing high accuracy compared to experiment and coupled cluster calculations on 12 excited states. 
The state-specific orbital optimization made possible by the new technique results in improvements in excited state energies from fixed node DMC.

\section{Penalty-based optimization using variational Monte Carlo}

The method solves for each energy eigenstate one at a time, by following the procedure outlined here:  
\begin{enumerate}
\item First, the stochastic reconfiguration method\cite{casulaCorrelatedGeminalWave2004} is used to find the VMC approximation to the ground state, $\ket{\Psi_0}$. 
\item The first excited state $\ket{\Psi_1}$ is found by optimizing the objective function Eqn~\ref{eqn:objective} with $\vec{S}^*=[0]$ and $\lambda$ set larger than the expected $E_1-E_0$.
The algorithm is not very sensitive to the value of $\lambda$, so we typically use $\lambda$ of order 1 Hartree, substantially higher than excitation energies in the systems considered here. 
\item The second excited state is found by optimizing the objective function Eqn~\ref{eqn:objective} with $\vec{S}^*=[0,0]$ and anchor states $\ket{\Psi_0}$ and $\ket{\Psi_1}$.
\item Further excited states are found in the same way, by orthogonalizing to the ones found in the previous steps.
\end{enumerate}

\subsection{Objective function}

\begin{figure}
\includegraphics{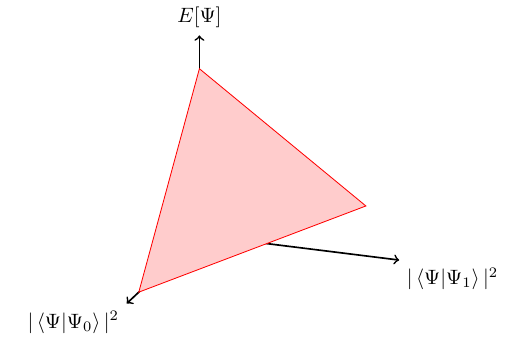}
    \caption{The lower bound of $E[\Psi]$ as a function of overlap with the first two eigenstates. The vertices are the first three eigenstates.}
\label{fig:le_space}
\end{figure}

As diagrammed in Fig~\ref{fig:le_space}, it is straightforward to show that if $\ket{\Phi_0}$ is the ground state of the Hamiltonian, then so long as $\lambda > E_1-E_0$ the function
\begin{equation}
    \argmin_\Psi ( E[\Psi] + \lambda N_\Psi^2 N_{\Phi_0}^2 |\braket{\Psi|\Phi_0}|^2 )
\end{equation}
is equal to the first excited state $\ket{\Phi_1}$, where $E[\Psi]$ is the expectation value of the energy and  $N_\Psi = 1/\sqrt{|\langle \Psi | \Psi \rangle|}$.

For completeness, we show this here. 
Consider the objective functional
\begin{align}
    O[\Psi] &= E[\Psi] +  N_\Psi^2 N_{\Phi_0}^2 \lambda_i |\langle \Psi | \Phi_0\rangle|^2 
\end{align}
Then the functional derivative of each term is given by 
\begin{align}
    \frac{\delta N_\Psi^2 N_{\Phi_0}^2 |\braket{\Psi | \Phi_0}|^2}{\delta \Psi^*} 
    =  N_\Psi^2 ( N_{\Phi_0}^2 \braket{\Phi_0 | \Psi} &\textcolor{blue}{\ket{\Phi_0}} \notag \\
    -  N_{\Psi}^2 N_{\Phi_0}^2|\braket{\Psi|\Phi_0}|^2&\textcolor{blue}{\ket{\Psi}})
\end{align}
and 
\begin{equation}
    \frac{\delta E[\Psi]}{\delta \Psi^*} = N_\Psi^2 (H - E[\Psi]) \textcolor{blue}{\ket{\Psi}}.
\end{equation}
We have highlighted the unpaired kets in blue for clarity.
Combining the two, and setting $\frac{\delta O}{\delta \Psi^*}$ to zero, we obtain
\begin{align}
    ( H -E[\Psi] - \lambda N_\Psi^2 N_{\Phi_0}^2 |\braket{\Psi|\Phi_0}|^2 ) & \textcolor{blue}{\ket{\Psi}} \notag \\
    + \lambda  N_{\Phi_0}^2 \braket{\Phi_0 | \Psi} & \textcolor{blue}{\ket{\Phi_0}} =0
\end{align}
The $H-E[\Psi]$ term ensures that this equation can only be satisfied by an an energy eigenstate $\ket{\Phi_i}$. 
For $i\neq 0$, $\ket{\Psi} = \ket{\Phi_i}$ is a solution because $\braket{\Phi_i | \Phi_0}=0$. 
$\ket{\Phi_0}$ is a solution since $\braket{\Phi_0 | \Phi_0}=1$, which allows the $\lambda$ terms to cancel.
The value of the functional $O[\Phi_i]=E_i + \lambda \delta_{i0}$. 
Therefore, if $\lambda > E_1-E_0$, the global minimum is at $\ket{\Phi_1}$. 
Because of the structure demonstrated in Fig~\ref{fig:le_space}, there are no local minima in the complete Hilbert space. 
Similarly, the $N$'th excited state is given by
\begin{equation}
    \ket{\Phi_N} = \argmin_\Psi 
    \left( E[\Psi] + \sum_i^{N-1} \lambda_i N_{\Psi}^2 N_{\Phi_0}^2|\braket{\Psi|\Phi_i}|^2 \right).
\end{equation}
as long as $\lambda_i > E_i - E_0$. 

We write the algorithm in terms of \textit{anchor states} $\ket{\Psi_i}$, where $i=0,\ldots,N-1$. 
These states are fixed during the optimization, and only the parameters of a single wave function $\ket{\Psi}$ are optimized.
While ideally the anchor states are energy eigenstates, in general they will be best approximations to them.
We also find it useful to consider the objective function 
\begin{equation}
    O[\Psi]= E[\Psi]+\sum_i \lambda_i |\vec{S_i}-\vec{S_i}^*|^2,
    \label{eqn:objective}
\end{equation}
where
\begin{equation}
    S_i = \frac{\braket{\Psi|\Psi_i}}{\sqrt{\braket{\Psi|\Psi}\braket{\Psi_i |\Psi_i}}},
\end{equation}
and $\vec{S}^*$ is a set of \textit{target} overlaps.
For example, to obtain the $N$'th excited state one would use $N-1$ anchor states each set to the best approximation to the $N-1$ lowest energy eigenstates and set $\vec{S}^*$ equal to a zero vector of $N-1$ length.

\begin{table}
    \caption{Implementation of the penalty-based optimization for excited states.}
    \label{table:algorithm}
    \begin{enumerate}
    \item Choose target overlaps $\vec{S}^*$
    \item Initialize $\vec{p_0}$
    \item \textbf{for} i in range(nsteps) \textbf{do}
     \begin{enumerate}
        
        \item Compute $N_i$, $\vec{S}$, $\nabla_p N_0$, $\nabla_p \vec{S}$, $E$, $\nabla_p E$
        \item \textbf{if} abs($N_0$-0.5) $>$ threshold, normalize $\Psi_N$
        \item Objective function is $O = E+\vec{\lambda}\cdot (\vec{S}-\vec{S}^*)^2$
        \item Construct gradient $\nabla_p O = \nabla_p E + \nabla_p (\vec{\lambda}\cdot(\vec{S}-\vec{S}^*)^2)$
        \item $\nabla_p N \leftarrow R^{-1} \nabla_p N$
        \item $\nabla_p O \leftarrow R^{-1} \nabla_p O$
        \item $\nabla_p O \leftarrow \nabla_p O - \frac{(\nabla_p O)\cdot(\nabla_p N)}{|\nabla_p N|^2} \nabla_p N$
        \item $\vec{p}(\tau) \leftarrow \vec{p}_{i-1} - \tau \nabla_p O$
        \item $\vec{p}_i \leftarrow \argmin_p (O(\vec{p}(\tau)),\tau)$
     \end{enumerate}
    \end{enumerate}
\end{table}

\subsection{Computation of the objective function and its derivatives using variational Monte Carlo}
\label{sec:computation}

In this section, we will explain how to evaluate the quantities needed using standard variational Monte Carlo techniques.\cite{foulkesQuantumMonteCarlo2001}
In this implementation, we sample a different distribution for each anchor state:
\begin{equation}
\rho_i(\bR)  = |\Psi_i(\bR)|^2  + |\Psi(\bR)|^2.
\label{eqn:rho}
\end{equation}
Then the unnormalized overlap is estimated in Monte Carlo
\begin{equation}
    \braket{\Psi_j|\Psi_k}_i \simeq \left\langle \frac{\Psi_j^*(\bR)\Psi_k(\bR)}{\rho_i (\bR)} \right\rangle_{\bR\sim\rho_i}
    \label{eqn:overlap},
\end{equation}
where $\bR$ is the many-electron coordinate and $\bR\sim \rho_i$ means that $\bR$ is sampled from the normalized distribution $\frac{\rho_i(\bR)}{\int \rho_i(\bR)}$.
Here the subscript $i$ indicates that the overlap was estimated using $\rho_i$.

The relative normalization of the wave function $\ket{\Psi}$ is thus
\begin{equation}
    N_i = \braket{\Psi|\Psi}_i,
\end{equation}
which is evaluated the same way as Eqn~\ref{eqn:overlap}.
For a parameter $p$ of $\ket{\Psi}$,    
\begin{equation}
\partial_p N_i = 2 \text{Re} \left\langle \frac{\partial_p \Psi^*(\bR)\Psi(\bR)}{\rho_i (\bR)} \right\rangle_{\bR\sim\rho_i}
\end{equation}

The overlap is given by 
\begin{equation}
    S_i = \frac{\braket{\Psi|\Psi_i}_i}{A_i},
\end{equation}
where $A_i=\sqrt{\braket{\Psi|\Psi}_i\braket{\Psi_i |\Psi_i}_i}$.
The derivative of the unnormalized overlap is
\begin{equation}
    \langle \partial_p \Psi | \Psi_i \rangle = \left\langle \frac{\partial_p \Psi^*(\bR)\Psi_i(\bR)}{\rho_i (\bR)} \right\rangle_{\bR\sim\rho_i}.
\end{equation}
The derivative of the normalized overlap is computed using the above components as follows
\begin{equation} 
    \partial_p S_{i} = \frac{\langle \partial_p \Psi | \Psi_i \rangle_i}{A_i} - \frac{1}{2}\frac{\langle \Psi | \Psi_i \rangle_i}{A_i} \frac{\partial_p N_i}{N_i}.
\end{equation}
Thus, all derivatives can be computed using only the wave function parameter derivatives.

We also regularize all derivatives using the stochastic reconfiguration\cite{casulaCorrelatedGeminalWave2004} step to compute
\begin{equation}
    R_{pq} = \braket{\partial_p \Psi | \partial_q \Psi },
    \label{eqn:sr_overlap}
\end{equation}
for parameter indices $p$ and $q$.

\subsection{Practical details}

The algorithm is outlined in Table~\ref{table:algorithm}. 
We provide comments on steps that have some nontrivial considerations.
In step \textbf{2}, the parameters are initialized. 
We typically initialize the parameters using an approximate excited state, typically either from an orbital promotion in a single determinant, or from a small quantum chemistry calculation.
We have checked that it is possible to optimize starting from the ground state, but such a strategy is unnecessarily expensive, in particular because the objective function of Eqn~\ref{eqn:objective} is a saddle point at the ground state.

In step \textbf{3(a)}, all the quantities in Sec~\ref{sec:computation} are computed. 
A Monte Carlo sampling of $\rho_i$ is done for each anchor wave function. 
The algorithm thus scales mildly with the number of anchor wave functions. 
The energy and its derivatives are averaged among all samples, so the most costly component of the calculation does not increase much with the number of anchor wave functions. 

It is important for the normalization of the wave functions to be similar; otherwise the density in Eqn~\ref{eqn:rho} is unbalanced. 
One could adjust weights in Eqn~\ref{eqn:rho}, but we find it more convenient to normalize all wave functions.
We ensure that all anchor states have the same normalization, and use the first anchor state as a reference.
Before performing an optimization move, we first check whether $N_0$ is too far from $\frac{1}{2}$. 
The threshold is typically about 0.1. 
If it is too far, the parameters are rescaled to normalize the wave function and VMC is repeated with the renormalized wave function parameters. 
This is performed in step \textbf{3(b)} in Table~\ref{table:algorithm}.

In steps \textbf{3(e)} and \textbf{3(f)}, we regularize the gradients of the normalization and the objective function. 
It is necessary to regularize both prior to the projection that will come in the next step. 

To prevent moves that change the normalization, we project out the derivative of the normalization from the objective function. 
Otherwise, the moves diverge from the equal normalization manifold, and it becomes difficult to evaluate the overlaps accurately.
This is performed in step \textbf{3(g)} in Table~\ref{table:algorithm}.\footnote{Since we are using multi-Slater-Jastrow wave functions in this work, it was more convenient to control the normalization of the wave function in this way.
Another option, not tried here, is to parameterize the wave function in a norm-conserving manner, which could increase the efficiency of the algorithm.
On the other hand, for emerging wave functions, it may be more convenient to allow the normalization to vary.}

We find that line minimization, performed in step \textbf{3(i)} in Table~\ref{table:algorithm}, improves the performance of the algorithm significantly.
We use correlated sampling to compute the objective function for various values of $\tau$, and fit to a quadratic. 
We also reject moves that change the relative normalization by more than 0.3.

\section{Demonstration of excited state optimization using VMC} 

To demonstrate the technique, we apply it to two cases; H$_2$ at varying bond lengths to check for correctness versus an exact solution, and the excited states of benzene to demonstrate that it is capable of optimizing about 10,000 parameters on a system with 30 electrons in the calculation. 

\subsection{Application to H$_2$}

\begin{figure}
    \includegraphics{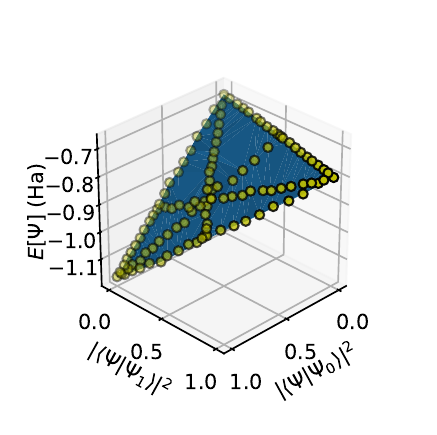}
    \caption{Targeted overlaps for H$_2$ in the space of the lowest three eigenstates. The yellow points are wave functions, using the penalty method to set $S^*$. The $x$ and $y$ coordinates are the measured overlaps after optimization, and the energy $E[\Psi]$ is the expectation value of the wave function after optimization. } 
    \label{fig:h2_3d_plane}
\end{figure}

For application to H$_2$, the trial wave function was a simple complete active space (CASCI) wave function with 2 electrons and 2 orbitals taken from restricted Hartree-Fock. 
We take the lowest three excited states from this calculation, labeled $\Psi_\textrm{CASCI,i}$, where $i$ runs from 0 to 2.
This wave function was modified using a 2-body Jastrow as parameterized in previous work\cite{WAGNER20093390} to obtain 
\begin{equation}
    \Psi_\textrm{CASCI-J,0} = \Psi_\textrm{CASCI,0} e^{U_0}
\end{equation}
We then optimized the determinant, orbital, and Jastrow parameters using a modified version of stochastic reconfiguration implemented in \texttt{pyqmc} to obtain $\Psi_\textrm{CASCI-J,0}$.

We then considered two Jastrow-based approximations to the excited states.
The first, which we denote ``Fixed," is commonly used in the literature. 
It is given by 
\begin{equation}
    \Psi_\textrm{Fixed CASCI-J, i} = \Psi_\textrm{CASCI,i} e^{U_0}.
    \label{eqn:fixed_casci}
\end{equation}
In this fixed wave function, no parameters are optimized at all; that is, the Jastrow is the same as the ground state and the determinant and orbital coefficients are kept fixed. 
The second, we denote ``Optimized," begins with Eqn~\ref{eqn:fixed_casci}, and uses the penalty method to optimize determinant, orbital, and Jastrow parameters while ensuring orthogonality to lower states. 
In the Supplementary Information, a Snakemake workflow\cite{kosterSnakemakeScalableBioinformatics2012} is provided that performs the calculations shown here in \texttt{pyqmc} and \texttt{pyscf}.\cite{doi:10.1002/wcms.1340}
We use a $\lambda$ of 2 Hartrees for these calculations, which were converged.

For reference values, we used full configuration interaction (Full CI) to compute exact energies in a finite basis.
We found that at the triple $\zeta$ level, the energies were fairly well converged. 
This is slightly earlier than most materials, due to the fact that H$_2$ is very simple. 

\begin{figure}
    \includegraphics{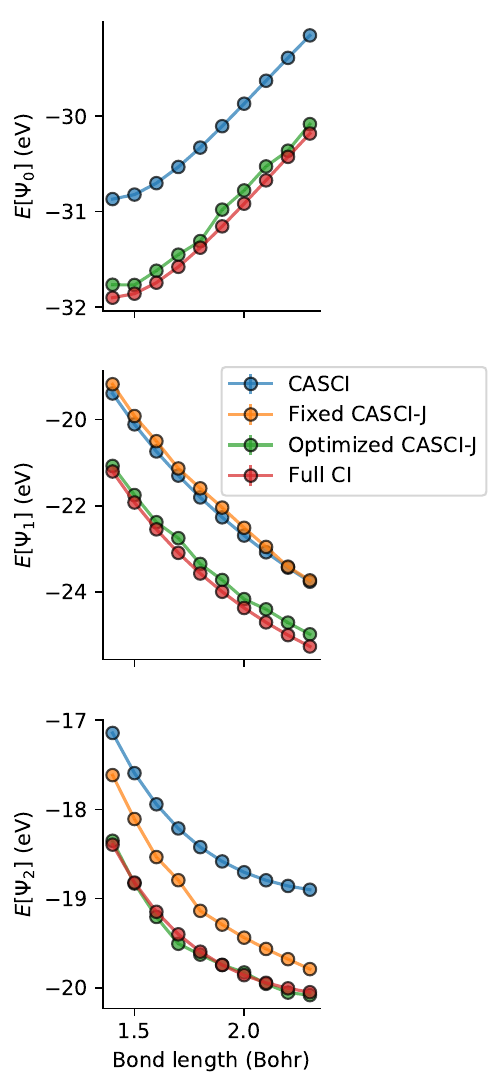}
    \caption{\label{fig:h2energy}
        Comparison of FCI eigenvalues with energies of wave functions optimized with orthogonal optimization to target the first three eigenstates for H$_2$.
       The cc-pvtz basis of Dunning\cite{dunningGaussianBasisSets1989} was used. 
    }
\end{figure}

In Fig~\ref{fig:h2_3d_plane}, we demonstrate the targeting capability of this technique. 
Each point is a wave function generated in the Dunning cc-pvdz\cite{dunningGaussianBasisSets1989} basis as described above, with $S^*$ set to various points on the Bloch sphere connecting the first 3 excited states. 
As expected, the superpositions of low energy wave functions fall on a plane as sketched in Fig~\ref{fig:le_space}, a critical check that the calculation is creating the desired wave functions.

Comparisons between the CASCI, QMC, and full CI results are presented in Fig~\ref{fig:h2energy}. 
With a rather compact wave function, the optimized CASCI-J wave function obtains close agreement with the exact calculation, while optimizing significantly from the starting fixed CASCI-J wave function. 
In the case of the first excited state, simply adding a Jastrow factor to an existing CASCI wave function does not improve the energies at all; optimization is required to obtain accurate results. 

\subsection{Application to benzene excited states}

\begin{figure}
    \includegraphics{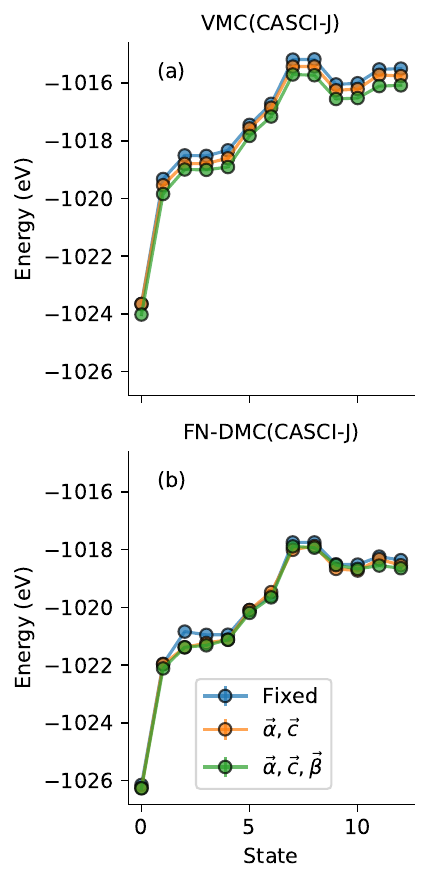}
    \caption{Benzene excited states computed using \textbf{(a)} VMC and \textbf{(b)} time-step extrapolated DMC for the full $\pi$-space spectrum. 
    Different colors refer to increased parameter sets. $\vec{\alpha}$ are the Jastrow coefficients, $\vec{c}$ are determinant coefficients, and $\vec{\beta}$ are orbital coefficients, as denoted in Eqn~\ref{eqn:benzene_wf} }
    \label{fig:benzene_orbopt}
\end{figure}

As a demonstration on a larger system, we computed the full $\pi$ space spectrum of benzene using our excited state optimization method.
This set of thirteen states contains rich physical structure, with two different spin channels, single and double electron excitations, and states with ionic bonding in contrast to the covalently bonded ground state.\cite{ROOS19925}
Some of these states, such as the $^1$E$_{1u}$ state, have been considered strongly correlated by previous authors.\cite{ROOS19925}
As such, these excitations are a standard benchmark set for excited state methods and have been used to validate and compare electronic structure techniques for some time.\cite{doi:10.1063/1.471985, doi:10.1063/1.2889385}

To represent the wave functions, we used a multi-Slater-Jastrow wave function parametrization of the form 
\begin{equation}
|\Psi(\vec{\alpha}, \vec{c}, \vec{\beta})\rangle = e^{J(\vec{\alpha})}\sum_{i} c_i |D_i (\vec{\beta})\rangle.
\label{eqn:benzene_wf}
\end{equation}
The Jastrow factor $J(\vec{\alpha})$ is a 2-body Jastrow\cite{WAGNER20093390} factor.
The determinants $|D_i\rangle$ in the multi-Slater expansion were selected from a minimal CASCI calculation over the six $\pi$ electrons and six $\pi$ orbitals in benzene.
The single particle orbitals used in the CASCI were computed using 
density functional theory (DFT) with the B3LYP functional, BFD triple-$\zeta$ basis, and BFD pseudo-potential.\cite{doi:10.1063/1.2741534}
The CASCI and DFT calculations were carried out using \texttt{pyscf}.\cite{doi:10.1002/wcms.1340}
The parameters in the wave function are: 108 Jastrow parameters $\vec{\alpha}$,  400 determinant coefficients $\vec{c}$, and 9,288 orbital parameters $\vec{\beta}$.
    
  \begin{table*}[ht]
    \caption{Comparison of theoretically computed excitation energies to experimental values. All values are in eV.
    Maximum indicates the transition of maximum intensity.
    The adiabatic and ZPVE corrections are estimated using TDDFT with the PBE functional and 6-31g basis.
	}
    \label{fig:benzene_table}
    \begin{tabular}{c|cc|cc|ccccc}
    \multicolumn{1}{c}{\textbf{}} & \multicolumn{2}{c}{\textbf{Spectroscopy}\cite{doi:10.1063/1.471985, doi:10.1063/1.1747242, doi:10.1063/1.435381, doi:10.1063/1.475801, BURSILL1998305}} & \multicolumn{2}{c}{\textbf{Corrections}} & \multicolumn{5}{c}{\textbf{Vertical excitation}}\\
  State  &  Maximum &  \textbf{$E_{00}$}  &  Adiabatic  &  ZPVE  &  Experiment  &  \textbf{DMC $\vec{\alpha}, \vec{c}, \vec{\beta}$}  &  \textbf{CC3}\cite{doi:10.1063/1.2889385}  &  \textbf{CASPT2}\cite{ROOS19925} &  \textbf{TDDFT-PBE0}\cite{doi:10.1063/1.479571}  \\
\hline
    $^1B_{2u}$    &        4.90        &        4.72        &        -0.19        &      -0.18      &            5.09           &                       5.15(3)                       &      5.08      &        4.7        &       5.39       \\
    $^1B_{1u}$    &        6.20        &        6.03        &       -0.19           &       -0.33          &    6.55             &                       6.62(4)                       &      6.54      &        6.1        &       6.05       \\
    $^1E_{1u}$    &        6.94        &        6.87        &       -0.24               &    -0.32             &  7.43                        &                       7.72(4)                       &      7.13      &       7.06        &       7.21       \\
    $^1E_{1u}$    &        6.94        &        6.87        &        -0.24              &    -0.32             &  7.43                        &                       7.63(3)                       &      7.13      &       7.06        &       7.21       \\
    $^1E_{2g}$    &      7.80(20)      &        7.81        &        -0.21              &    -0.45             &  8.47                        &                       8.38(3)                       &      8.41      &       7.77        &       7.52       \\
    $^1E_{2g}$    &      7.80(20)      &        7.81        &        -0.21              &    -0.45             &  8.47                        &                       8.34(3)                       &      8.41      &       7.77        &       7.52       \\
    $^3B_{1u}$    &        3.94        &        3.65        &        -0.55        &     -0.19   &            4.39           &                       4.15(3)                       &      4.15      &       3.94        &       3.82       \\
    $^3E_{1u}$    &        4.76        &        4.63        &                      &                 &                          &                       4.89(3)                       &      4.86      &        4.5        &       4.7        \\
    $^3E_{1u}$    &        4.76        &        4.63        &                      &                 &                          &                       4.96(4)                       &      4.86      &        4.5        &       4.7        \\
    $^3B_{2u}$    &        5.60        &        5.58        &                      &                 &                          &                       6.08(4)                       &      5.88      &       5.44        &       5.05       \\
    $^3E_{2g}$    &      7.49(25)      &      7.49(25)      &                      &                 &                          &                       7.74(4)                       &      7.51      &       7.03        &       7.18       \\
    $^3E_{2g}$    &      7.49(25)      &      7.49(25)      &                      &                 &                          &                       7.60(4)                       &      7.51      &       7.03        &       7.18       \\
\hline
\end{tabular}

\end{table*}

We used the parameterization of Eqn~\ref{eqn:benzene_wf} to compute the benzene spectra using three different methods, with increasing cost.
The first method, denoted fixed, is a standard QMC excited state technique, where the coefficients $\vec{\alpha}, \vec{c}$ are first optimized on the ground state CASCI root with frozen orbital coefficients $\vec{\beta}$, then the optimized Jastrow is multiplied with higher energy CASCI roots, and finally these trial wave functions are used in VMC to compute excited state energies.
This method does not allow for state-specific optimization of any of the parameters $\vec{\alpha}, \vec{c},$ or $\vec{\beta}$.

To understand the effects of orbital optimization in this system, we consider two parameter sets using the penalty technique. 
In the first, we fix the orbital parameters $\vec{\beta}$ to the DFT ground state orbital coefficients, but allow the other parameters $\vec{\alpha}, \vec{c}$ to be optimized;  we denote the wave functions as VMC $\vec{\alpha}, \vec{c}$.
Finally, we optimize all coefficients in Eqn~\ref{eqn:benzene_wf}, denoting those wave functions $\vec{\alpha}, \vec{c},\vec{\beta}$. 
All QMC calculations were carried out in \texttt{pyqmc}.\cite{pyqmc}

The results of our excited state computations at the VMC level are shown in Fig~\ref{fig:benzene_orbopt}a.
We see a consistent 0.2 eV decrease in total energy across all twelve excited states going from the fixed parameter QMC method to the $\vec{\alpha}, \vec{c}$ method using our new optimization technique.
We find that optimizing $\vec{\alpha}, \vec{c}, \vec{\beta}$ yields up to a 0.5 eV reduction in total energy relative to the fixed technique across nearly all of the excited states, a gain of around 0.3 eV due to orbital optimization.
However, the differences between the excited states and the ground state are very similar between the frozen and optimized orbital calculations, which demonstrates the extent to which standard excited state techniques depend heavily on error cancellation.

We computed the time-step extrapolated DMC energies for the computed excited state wave functions, which are shown in Fig ~\ref{fig:benzene_orbopt}b.
The additional benefit of optimizing the orbitals in DMC is seen primarily for just three states, states 1, 6, and 11, with reductions in energy of 0.14(3), 0.18(3), 0.22(3) eV respectively.
The latter two of these states are the $^1 B_{1u}, ^1 E_{1u}$ states which have strong ionic bonding character among the $\pi$ orbitals, unlike the ground state which has covalent bonding character.\cite{ROOS19925}
This difference in bonding character is captured by the orbital optimization, leading to larger reductions in total energy in these states, while the other states have no reduction in total energy.
The fact that only some excited states benefit from orbital optimization in DMC means that the energy differences are affected; ultimately they are improved.

Before comparing to experiment, we note that most theoretical results report the vertical excitation energy, which omits nuclear relaxation and vibrational effects. 
The transition from the ground vibrational level of the ground state to the ground vibrational level of the excited state is called the 0-0 excitation energy ($E_{00}$), and is not necessarily the maximum intensity. 
The vertical excitation energy can be related to the 0-0 excitation energy as follows:
\begin{align}
E_{00}(0 \rightarrow j) &=  E_j(\vec{X}_0) - E_0(\vec{X}_0)  \textrm{ (vertical)} \notag \\
                           & + E_j(\vec{X}_j) - E_j(\vec{X}_0)  \textrm{ (adiabatic)} \notag \\
                           & + \frac{\hbar}{2} \sum_{i} (\omega_{j, i} - \omega_{0, i}). \textrm{ (ZPVE)},
\label{eq:energy_decomp}
\end{align}
where $E_j$ is the Born-Oppenheimer energy of state $j$, $X_j$ are the set of nuclear coordinates at the minimum energy of state $j$, and $\omega_{j,i}$ is the frequency of vibrational mode $i$ for electronic state $j$.
The first line is the fixed nuclei vertical excitation energy, the quantity which is directly computed in our work and the cited theoretical calculations, where the total energy of the $j$th excited state is computed at the ground state equilibrium nuclear positions $\vec{X}_0$.
The second line is the adiabatic correction, which accounts for shifting of the nuclear positions in the $j$th excited state to the configuration $\vec{X}_j$.
The last line is the zero-point vibrational energy (ZPVE) correction, and accounts for changes in the vibrational degrees of freedom between the ground and excited state.
The experimental vertical excitation is obtained by subtracting the adiabatic and ZPVE energies from the 0-0 excitation energies. 
 
In Table~\ref{fig:benzene_table}, we compare the vertical excitation energy computed from experiment, the fully optimized DMC results in this work, and literature results of large-basis coupled cluster (CC3)\cite{doi:10.1063/1.2889385}, complete active space perturbation theory(CASPT2)\cite{ROOS19925}, and time dependent density functional theory using the PBE0 functional (TDDFT-PBE0)\cite{doi:10.1063/1.479571}.
We also report computed ZPVE and 0-0 corrections using TDDFT with a 6-31g basis and PBE functional for the singlet states; we encountered difficulties converging the triplet states in TDDFT. 
We believe the TDDFT estimates to be reasonable given the agreement with computed and measured values in literature: -0.15 eV \cite{doi:10.1063/1.475801} and -0.17 eV \cite{Ziegler1982TheVS} adiabatic and ZPVE corrections for the $^1 B_{2u}$ state, -0.56 eV \cite{doi:10.1063/1.3270190} and -0.17 eV \cite{OHNO2002409} for the $^3 B_{1u}$ state.
For the available vertical excitations, the DMC and CC3 results both are within 0.2 eV of the experimental values, while the CASPT2 and TDDFT method have larger differences. 
The corrections due to adiabatic and ZPVE effects are large enough that this conclusion might be reversed without these effects included. 

\begin{table}
\caption{Agreement between theoretical vertical excitation energies. Optimizing more parameters using the penalty method improves the agreement between CC3 and DMC, as seen in the mean and standard deviation of the difference between the excitation energies. }
\label{table:mean_comparison}
\begin{tabular}{llcc}
\hline
          Method &  Parameters & \multicolumn{2}{c}{$\Delta E_j = E_j($m$)-E_j($CC3$)$} \\
        & &   mean (eV) &    RMS (eV) \\
\hline
    CASPT2 &                                  & -0.38 &  0.42 \\
    TTDFT-PBE0 &                              & -0.33 &  0.50 \\
    VMC &       None                          &  0.34 &  0.46 \\
    VMC & $\vec{\alpha},\vec{c}$              &  0.12 &  0.35 \\
    VMC & $\vec{\alpha},\vec{c}, \vec{\beta}$ &  0.19 &  0.35 \\    
    DMC &       None                          &  0.24 &  0.35 \\
    DMC & $\vec{\alpha},\vec{c}$              &  0.18 &  0.31 \\
    DMC & $\vec{\alpha},\vec{c}, \vec{\beta}$ &  0.15 &  0.24 \\
\hline
\end{tabular}
\end{table}

In Table~\ref{table:mean_comparison}, we list the mean and standard deviation of the difference between CC3 and other methods. 
CC3 has been found to be very accurate for electronic excitations, including correlated doubles-like excitations for benzene.\cite{ROOS19925,loosReferenceEnergiesDouble2019}
The consistency between accurate, explicitly correlated methods like CC3 and our new QMC technique, which make different approximations, is encouraging, as is the fact that the QMC results approach the CC3 results as the number of optimized parameters is increased.

\begin{figure}
    \includegraphics{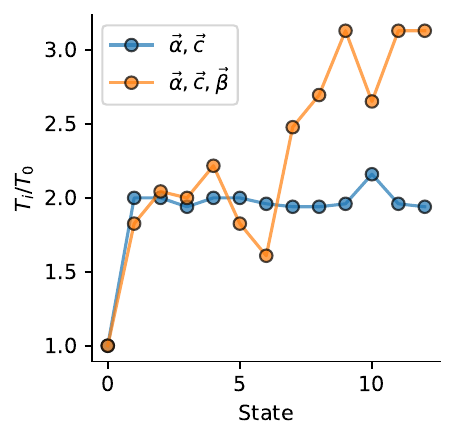}
    \caption{Relative cost $T_i/T_0$ of computing excited states in VMC for the two different parameterizations considered on the benzene molecule, where $T_i$ is the CPU time required to perform the calculation of excited state $i$.
    $T_0$ for the ground state were 1.25 and 23.00 hours on a single 40 processor node for the $\vec{\alpha}, \vec{c}$ and $\vec{\alpha}, \vec{c}, \vec{\beta}$ parameterizations respectively.
	}
    \label{fig:benzene_cost}
    \end{figure}

The cost of each excited state calculation is roughly a factor of 2-3 higher than the ground state calculation, as shown in Fig~\ref{fig:benzene_cost}. 
As a point of reference, the ground state optimization took 1.25 hours on a 40 core processor for the fixed orbital optimization and 23 hours on a modern 40 core processor for the orbital optimized calculation. 
One should keep in mind that as is typical in Monte Carlo, this cost is highly dependent on the desired uncertainty in the result.
We chose very converged parameters, which results in a relatively high computational cost.
The total relative cost over the ground state for all 13 states is a factor of 25 for the $\vec{\alpha}, \vec{c}$ calculation and 30 for $\vec{\alpha}, \vec{c}, \vec{\beta}$.
Importantly, the relative cost does not increase rapidly with the state, since as mentioned in the introduction the overlaps are not very expensive to evaluate.

\section{Conclusion}
We have presented a scalable algorithm to compute approximate excited states of many-body systems using variational Monte Carlo. 
Our method is somewhat less complicated to implement than the linear method of Filippi\cite{doi:10.1021/ct900227j, doi:10.1021/acs.jctc.9b00476, doi:10.1021/ct1006295}, since the derivatives of the local energy are not required.
Further, our technique is capable of optimizing complex parameters like orbital coefficients in a state-specific manner, and likely would be able to optimize parameters from wave functions such as backflow\cite{kwonEffectsThreebodyBackflow1993,lopezriosInhomogeneousBackflowTransformations2006a} and neural network forms,\cite{PhysRevB.78.041101, PhysRevLett.122.226401,pfauInitioSolutionManyelectron2020,carleoSolvingQuantumManybody2017}  as well as pairing functions\cite{PhysRevB.77.115112}, since the complexity is not much higher than normal wave function optimization. 
In comparison to the method of Neuscamman \cite{doi:10.1021/acs.jctc.6b00508, doi:10.1021/acs.jctc.7b00923, doi:10.1021/acs.jpca.8b10671, doi:10.1021/acs.jctc.8b00879}, this technique does not require a tuned parameter $\omega$.
One positive aspect of this is that the penalty technique can access several distinct but degenerate excited states separately. 
Degenerate ground states would also be detected by the penalty technique.

The penalty technique is capable of optimizing wave functions with any overlap with a reference state. 
This capability may be useful in some circumstances, particularly for strongly correlated systems and magnetic systems, in which energy eigenstates may not be easily representable by simple wave functions, but non-orthogonal basis states can be used to make relevant experimental predictions.\cite{PhysRevLett.114.176401}
Such wave functions are also appropriate for usage in density matrix downfolding.\cite{10.3389/fphy.2018.00043}

The study on the benzene molecule revealed a few interesting physical insights. 
First, orbital optimization improves the nodal surface for excited states, particularly those with significant ionic character as compared to the ground state. 
Secondly, the classical association of the vertical excitation with the maximum intensity in spectroscopy leads to errors up to 0.4 eV in the excitation energies of benzene. 
A fully quantum treatment of the Franck-Condon principle, as shown here, brings the experimental estimates in closer alignment with coupled cluster and quantum Monte Carlo results.

\section{Supplementary material}

The supplementary material includes all collected data presented in the figures of the graph in comma separated value format. 
It also includes scripts and a workflow for the H$_2$ calculation for reproducibility.

\section{Acknowledgements} 

We would like to thank David Ceperley for helpful comments regarding the theorem of MacDonald\cite{macdonaldSuccessiveApproximationsRayleighRitz1933}, and William Wheeler for a careful reading of the code and bugfixes. 

This material is partially based upon work supported by the U.S. Department of Energy, Office of Science, Office of Basic Energy Sciences,  Computational Materials Sciences  program under Award Number DE-SC-0020177.
B.B. was supported by the Flatiron Institute.
The Flatiron Institute is a division of the Simons Foundation.
L.K.W. was supported by the Simons Collaboration on the Many-Electron Problem. 
This research is part of the Blue Waters sustained-petascale computing project, which is supported by the National Science Foundation (awards OCI-0725070 and ACI-1238993) the State of Illinois, and as of December, 2019, the National Geospatial-Intelligence Agency. Blue Waters is a joint effort of the University of Illinois at Urbana-Champaign and its National Center for Supercomputing Applications.

\section{Data availability}

The data that supports the findings of this study are available within the article and its supplementary material. 
A script and workflow in Snakemake\cite{kosterSnakemakeScalableBioinformatics2012} completely document the procedure used to perform the H$_2$ calculations.

\bibliography{ref}

\begin{thebibliography}{56}%
\makeatletter
\providecommand \@ifxundefined [1]{%
 \@ifx{#1\undefined}
}%
\providecommand \@ifnum [1]{%
 \ifnum #1\expandafter \@firstoftwo
 \else \expandafter \@secondoftwo
 \fi
}%
\providecommand \@ifx [1]{%
 \ifx #1\expandafter \@firstoftwo
 \else \expandafter \@secondoftwo
 \fi
}%
\providecommand \natexlab [1]{#1}%
\providecommand \enquote  [1]{``#1''}%
\providecommand \bibnamefont  [1]{#1}%
\providecommand \bibfnamefont [1]{#1}%
\providecommand \citenamefont [1]{#1}%
\providecommand \href@noop [0]{\@secondoftwo}%
\providecommand \href [0]{\begingroup \@sanitize@url \@href}%
\providecommand \@href[1]{\@@startlink{#1}\@@href}%
\providecommand \@@href[1]{\endgroup#1\@@endlink}%
\providecommand \@sanitize@url [0]{\catcode `\\12\catcode `\$12\catcode
  `\&12\catcode `\#12\catcode `\^12\catcode `\_12\catcode `\%12\relax}%
\providecommand \@@startlink[1]{}%
\providecommand \@@endlink[0]{}%
\providecommand \url  [0]{\begingroup\@sanitize@url \@url }%
\providecommand \@url [1]{\endgroup\@href {#1}{\urlprefix }}%
\providecommand \urlprefix  [0]{URL }%
\providecommand \Eprint [0]{\href }%
\providecommand \doibase [0]{http://dx.doi.org/}%
\providecommand \selectlanguage [0]{\@gobble}%
\providecommand \bibinfo  [0]{\@secondoftwo}%
\providecommand \bibfield  [0]{\@secondoftwo}%
\providecommand \translation [1]{[#1]}%
\providecommand \BibitemOpen [0]{}%
\providecommand \bibitemStop [0]{}%
\providecommand \bibitemNoStop [0]{.\EOS\space}%
\providecommand \EOS [0]{\spacefactor3000\relax}%
\providecommand \BibitemShut  [1]{\csname bibitem#1\endcsname}%
\let\auto@bib@innerbib\@empty
\bibitem [{\citenamefont {{Simons Collaboration on the Many-Electron Problem}}\
  \emph {et~al.}(2015)\citenamefont {{Simons Collaboration on the Many-Electron
  Problem}}, \citenamefont {LeBlanc}, \citenamefont {Antipov}, \citenamefont
  {Becca}, \citenamefont {Bulik}, \citenamefont {Chan}, \citenamefont {Chung},
  \citenamefont {Deng}, \citenamefont {Ferrero}, \citenamefont {Henderson},
  \citenamefont {Jiménez-Hoyos}, \citenamefont {Kozik}, \citenamefont {Liu},
  \citenamefont {Millis}, \citenamefont {Prokof’ev}, \citenamefont {Qin},
  \citenamefont {Scuseria}, \citenamefont {Shi}, \citenamefont {Svistunov},
  \citenamefont {Tocchio}, \citenamefont {Tupitsyn}, \citenamefont {White},
  \citenamefont {Zhang}, \citenamefont {Zheng}, \citenamefont {Zhu},\ and\
  \citenamefont
  {Gull}}]{simonscollaborationonthemany-electronproblemSolutionsTwoDimensionalHubbard2015}%
  \BibitemOpen
  \bibfield  {author} {\bibinfo {author} {\bibnamefont {{Simons Collaboration
  on the Many-Electron Problem}}}, \bibinfo {author} {\bibfnamefont
  {J.}~\bibnamefont {LeBlanc}}, \bibinfo {author} {\bibfnamefont {A.~E.}\
  \bibnamefont {Antipov}}, \bibinfo {author} {\bibfnamefont {F.}~\bibnamefont
  {Becca}}, \bibinfo {author} {\bibfnamefont {I.~W.}\ \bibnamefont {Bulik}},
  \bibinfo {author} {\bibfnamefont {G.~K.-L.}\ \bibnamefont {Chan}}, \bibinfo
  {author} {\bibfnamefont {C.-M.}\ \bibnamefont {Chung}}, \bibinfo {author}
  {\bibfnamefont {Y.}~\bibnamefont {Deng}}, \bibinfo {author} {\bibfnamefont
  {M.}~\bibnamefont {Ferrero}}, \bibinfo {author} {\bibfnamefont {T.~M.}\
  \bibnamefont {Henderson}}, \bibinfo {author} {\bibfnamefont {C.~A.}\
  \bibnamefont {Jiménez-Hoyos}}, \bibinfo {author} {\bibfnamefont
  {E.}~\bibnamefont {Kozik}}, \bibinfo {author} {\bibfnamefont {X.-W.}\
  \bibnamefont {Liu}}, \bibinfo {author} {\bibfnamefont {A.~J.}\ \bibnamefont
  {Millis}}, \bibinfo {author} {\bibfnamefont {N.}~\bibnamefont {Prokof’ev}},
  \bibinfo {author} {\bibfnamefont {M.}~\bibnamefont {Qin}}, \bibinfo {author}
  {\bibfnamefont {G.~E.}\ \bibnamefont {Scuseria}}, \bibinfo {author}
  {\bibfnamefont {H.}~\bibnamefont {Shi}}, \bibinfo {author} {\bibfnamefont
  {B.}~\bibnamefont {Svistunov}}, \bibinfo {author} {\bibfnamefont {L.~F.}\
  \bibnamefont {Tocchio}}, \bibinfo {author} {\bibfnamefont {I.}~\bibnamefont
  {Tupitsyn}}, \bibinfo {author} {\bibfnamefont {S.~R.}\ \bibnamefont {White}},
  \bibinfo {author} {\bibfnamefont {S.}~\bibnamefont {Zhang}}, \bibinfo
  {author} {\bibfnamefont {B.-X.}\ \bibnamefont {Zheng}}, \bibinfo {author}
  {\bibfnamefont {Z.}~\bibnamefont {Zhu}}, \ and\ \bibinfo {author}
  {\bibfnamefont {E.}~\bibnamefont {Gull}},\ }\href {\doibase
  10.1103/PhysRevX.5.041041} {\bibfield  {journal} {\bibinfo  {journal}
  {Physical Review X}\ }\textbf {\bibinfo {volume} {5}},\ \bibinfo {pages}
  {041041} (\bibinfo {year} {2015})}\BibitemShut {NoStop}%
\bibitem [{\citenamefont {Foulkes}\ \emph {et~al.}(2001)\citenamefont
  {Foulkes}, \citenamefont {Mitas}, \citenamefont {Needs},\ and\ \citenamefont
  {Rajagopal}}]{foulkesQuantumMonteCarlo2001}%
  \BibitemOpen
  \bibfield  {author} {\bibinfo {author} {\bibfnamefont {W.~M.~C.}\
  \bibnamefont {Foulkes}}, \bibinfo {author} {\bibfnamefont {L.}~\bibnamefont
  {Mitas}}, \bibinfo {author} {\bibfnamefont {R.~J.}\ \bibnamefont {Needs}}, \
  and\ \bibinfo {author} {\bibfnamefont {G.}~\bibnamefont {Rajagopal}},\ }\href
  {\doibase 10.1103/RevModPhys.73.33} {\bibfield  {journal} {\bibinfo
  {journal} {Reviews of Modern Physics}\ }\textbf {\bibinfo {volume} {73}},\
  \bibinfo {pages} {33} (\bibinfo {year} {2001})}\BibitemShut {NoStop}%
\bibitem [{\citenamefont {{Simons Collaboration on the Many-Electron Problem}}\
  \emph {et~al.}(2020)\citenamefont {{Simons Collaboration on the Many-Electron
  Problem}}, \citenamefont {Williams}, \citenamefont {Yao}, \citenamefont {Li},
  \citenamefont {Chen}, \citenamefont {Shi}, \citenamefont {Motta},
  \citenamefont {Niu}, \citenamefont {Ray}, \citenamefont {Guo}, \citenamefont
  {Anderson}, \citenamefont {Li}, \citenamefont {Tran}, \citenamefont {Yeh},
  \citenamefont {Mussard}, \citenamefont {Sharma}, \citenamefont {Bruneval},
  \citenamefont {van Schilfgaarde}, \citenamefont {Booth}, \citenamefont
  {Chan}, \citenamefont {Zhang}, \citenamefont {Gull}, \citenamefont {Zgid},
  \citenamefont {Millis}, \citenamefont {Umrigar},\ and\ \citenamefont
  {Wagner}}]{simonscollaborationonthemany-electronproblemDirectComparisonManyBody2020}%
  \BibitemOpen
  \bibfield  {author} {\bibinfo {author} {\bibnamefont {{Simons Collaboration
  on the Many-Electron Problem}}}, \bibinfo {author} {\bibfnamefont {K.~T.}\
  \bibnamefont {Williams}}, \bibinfo {author} {\bibfnamefont {Y.}~\bibnamefont
  {Yao}}, \bibinfo {author} {\bibfnamefont {J.}~\bibnamefont {Li}}, \bibinfo
  {author} {\bibfnamefont {L.}~\bibnamefont {Chen}}, \bibinfo {author}
  {\bibfnamefont {H.}~\bibnamefont {Shi}}, \bibinfo {author} {\bibfnamefont
  {M.}~\bibnamefont {Motta}}, \bibinfo {author} {\bibfnamefont
  {C.}~\bibnamefont {Niu}}, \bibinfo {author} {\bibfnamefont {U.}~\bibnamefont
  {Ray}}, \bibinfo {author} {\bibfnamefont {S.}~\bibnamefont {Guo}}, \bibinfo
  {author} {\bibfnamefont {R.~J.}\ \bibnamefont {Anderson}}, \bibinfo {author}
  {\bibfnamefont {J.}~\bibnamefont {Li}}, \bibinfo {author} {\bibfnamefont
  {L.~N.}\ \bibnamefont {Tran}}, \bibinfo {author} {\bibfnamefont {C.-N.}\
  \bibnamefont {Yeh}}, \bibinfo {author} {\bibfnamefont {B.}~\bibnamefont
  {Mussard}}, \bibinfo {author} {\bibfnamefont {S.}~\bibnamefont {Sharma}},
  \bibinfo {author} {\bibfnamefont {F.}~\bibnamefont {Bruneval}}, \bibinfo
  {author} {\bibfnamefont {M.}~\bibnamefont {van Schilfgaarde}}, \bibinfo
  {author} {\bibfnamefont {G.~H.}\ \bibnamefont {Booth}}, \bibinfo {author}
  {\bibfnamefont {G.~K.-L.}\ \bibnamefont {Chan}}, \bibinfo {author}
  {\bibfnamefont {S.}~\bibnamefont {Zhang}}, \bibinfo {author} {\bibfnamefont
  {E.}~\bibnamefont {Gull}}, \bibinfo {author} {\bibfnamefont {D.}~\bibnamefont
  {Zgid}}, \bibinfo {author} {\bibfnamefont {A.}~\bibnamefont {Millis}},
  \bibinfo {author} {\bibfnamefont {C.~J.}\ \bibnamefont {Umrigar}}, \ and\
  \bibinfo {author} {\bibfnamefont {L.~K.}\ \bibnamefont {Wagner}},\ }\href
  {\doibase 10.1103/PhysRevX.10.011041} {\bibfield  {journal} {\bibinfo
  {journal} {Physical Review X}\ }\textbf {\bibinfo {volume} {10}},\ \bibinfo
  {pages} {011041} (\bibinfo {year} {2020})}\BibitemShut {NoStop}%
\bibitem [{\citenamefont {Jastrow}(1955)}]{jastrowManyBodyProblemStrong1955}%
  \BibitemOpen
  \bibfield  {author} {\bibinfo {author} {\bibfnamefont {R.}~\bibnamefont
  {Jastrow}},\ }\href {\doibase 10.1103/PhysRev.98.1479} {\bibfield  {journal}
  {\bibinfo  {journal} {Physical Review}\ }\textbf {\bibinfo {volume} {98}},\
  \bibinfo {pages} {1479} (\bibinfo {year} {1955})}\BibitemShut {NoStop}%
\bibitem [{\citenamefont {Ma}\ \emph {et~al.}(2013)\citenamefont {Ma},
  \citenamefont {Zhang},\ and\ \citenamefont
  {Krakauer}}]{maExcitedStateCalculations2013}%
  \BibitemOpen
  \bibfield  {author} {\bibinfo {author} {\bibfnamefont {F.}~\bibnamefont
  {Ma}}, \bibinfo {author} {\bibfnamefont {S.}~\bibnamefont {Zhang}}, \ and\
  \bibinfo {author} {\bibfnamefont {H.}~\bibnamefont {Krakauer}},\ }\href
  {\doibase 10.1088/1367-2630/15/9/093017} {\bibfield  {journal} {\bibinfo
  {journal} {New Journal of Physics}\ }\textbf {\bibinfo {volume} {15}},\
  \bibinfo {pages} {093017} (\bibinfo {year} {2013})}\BibitemShut {NoStop}%
\bibitem [{\citenamefont {{Mitas}}\ and\ \citenamefont
  {Martin}(1994)}]{mitasQuantumMonteCarlo1994a}%
  \BibitemOpen
  \bibfield  {author} {\bibinfo {author} {\bibfnamefont {L.}~\bibnamefont
  {{Mitas}}}\ and\ \bibinfo {author} {\bibfnamefont {R.~M.}\ \bibnamefont
  {Martin}},\ }\href {\doibase 10.1103/PhysRevLett.72.2438} {\bibfield
  {journal} {\bibinfo  {journal} {Physical Review Letters}\ }\textbf {\bibinfo
  {volume} {72}},\ \bibinfo {pages} {2438} (\bibinfo {year}
  {1994})}\BibitemShut {NoStop}%
\bibitem [{\citenamefont {Williamson}\ \emph {et~al.}(1998)\citenamefont
  {Williamson}, \citenamefont {Hood}, \citenamefont {Needs},\ and\
  \citenamefont {Rajagopal}}]{williamsonDiffusionQuantumMonte1998}%
  \BibitemOpen
  \bibfield  {author} {\bibinfo {author} {\bibfnamefont {A.~J.}\ \bibnamefont
  {Williamson}}, \bibinfo {author} {\bibfnamefont {R.~Q.}\ \bibnamefont
  {Hood}}, \bibinfo {author} {\bibfnamefont {R.~J.}\ \bibnamefont {Needs}}, \
  and\ \bibinfo {author} {\bibfnamefont {G.}~\bibnamefont {Rajagopal}},\ }\href
  {\doibase 10.1103/PhysRevB.57.12140} {\bibfield  {journal} {\bibinfo
  {journal} {Physical Review B}\ }\textbf {\bibinfo {volume} {57}},\ \bibinfo
  {pages} {12140} (\bibinfo {year} {1998})}\BibitemShut {NoStop}%
\bibitem [{\citenamefont {Schiller}\ \emph {et~al.}(2015)\citenamefont
  {Schiller}, \citenamefont {Wagner},\ and\ \citenamefont
  {Ertekin}}]{schillerPhaseStabilityProperties2015}%
  \BibitemOpen
  \bibfield  {author} {\bibinfo {author} {\bibfnamefont {J.~A.}\ \bibnamefont
  {Schiller}}, \bibinfo {author} {\bibfnamefont {L.~K.}\ \bibnamefont
  {Wagner}}, \ and\ \bibinfo {author} {\bibfnamefont {E.}~\bibnamefont
  {Ertekin}},\ }\href {\doibase 10.1103/PhysRevB.92.235209} {\bibfield
  {journal} {\bibinfo  {journal} {Physical Review B}\ }\textbf {\bibinfo
  {volume} {92}},\ \bibinfo {pages} {235209} (\bibinfo {year}
  {2015})}\BibitemShut {NoStop}%
\bibitem [{\citenamefont {Zheng}\ and\ \citenamefont
  {Wagner}(2015{\natexlab{a}})}]{zhengComputationCorrelatedMetalInsulator2015}%
  \BibitemOpen
  \bibfield  {author} {\bibinfo {author} {\bibfnamefont {H.}~\bibnamefont
  {Zheng}}\ and\ \bibinfo {author} {\bibfnamefont {L.~K.}\ \bibnamefont
  {Wagner}},\ }\href {\doibase 10.1103/PhysRevLett.114.176401} {\bibfield
  {journal} {\bibinfo  {journal} {Physical Review Letters}\ }\textbf {\bibinfo
  {volume} {114}},\ \bibinfo {pages} {176401} (\bibinfo {year}
  {2015}{\natexlab{a}})}\BibitemShut {NoStop}%
\bibitem [{\citenamefont {Wagner}\ and\ \citenamefont
  {Ceperley}(2016)}]{wagnerDiscoveringCorrelatedFermions2016}%
  \BibitemOpen
  \bibfield  {author} {\bibinfo {author} {\bibfnamefont {L.~K.}\ \bibnamefont
  {Wagner}}\ and\ \bibinfo {author} {\bibfnamefont {D.~M.}\ \bibnamefont
  {Ceperley}},\ }\href {\doibase 10.1088/0034-4885/79/9/094501} {\bibfield
  {journal} {\bibinfo  {journal} {Reports on Progress in Physics}\ }\textbf
  {\bibinfo {volume} {79}},\ \bibinfo {pages} {094501} (\bibinfo {year}
  {2016})}\BibitemShut {NoStop}%
\bibitem [{\citenamefont {Frank}\ \emph {et~al.}(2019)\citenamefont {Frank},
  \citenamefont {Derian}, \citenamefont {Tokár}, \citenamefont {Mitas},
  \citenamefont {Fabian},\ and\ \citenamefont
  {Štich}}]{frankManyBodyQuantumMonte2019}%
  \BibitemOpen
  \bibfield  {author} {\bibinfo {author} {\bibfnamefont {T.}~\bibnamefont
  {Frank}}, \bibinfo {author} {\bibfnamefont {R.}~\bibnamefont {Derian}},
  \bibinfo {author} {\bibfnamefont {K.}~\bibnamefont {Tokár}}, \bibinfo
  {author} {\bibfnamefont {L.}~\bibnamefont {Mitas}}, \bibinfo {author}
  {\bibfnamefont {J.}~\bibnamefont {Fabian}}, \ and\ \bibinfo {author}
  {\bibfnamefont {I.}~\bibnamefont {Štich}},\ }\href {\doibase
  10.1103/PhysRevX.9.011018} {\bibfield  {journal} {\bibinfo  {journal}
  {Physical Review X}\ }\textbf {\bibinfo {volume} {9}},\ \bibinfo {pages}
  {011018} (\bibinfo {year} {2019})}\BibitemShut {NoStop}%
\bibitem [{\citenamefont {Wang}\ \emph {et~al.}(2020)\citenamefont {Wang},
  \citenamefont {Annaberdiyev},\ and\ \citenamefont
  {Mitas}}]{wangBindingExcitationsSi2020}%
  \BibitemOpen
  \bibfield  {author} {\bibinfo {author} {\bibfnamefont {G.}~\bibnamefont
  {Wang}}, \bibinfo {author} {\bibfnamefont {A.}~\bibnamefont {Annaberdiyev}},
  \ and\ \bibinfo {author} {\bibfnamefont {L.}~\bibnamefont {Mitas}},\ }\href
  {http://arxiv.org/abs/2007.11139} {\bibfield  {journal} {\bibinfo  {journal}
  {arXiv:2007.11139 [cond-mat, physics:physics]}\ } (\bibinfo {year} {2020})},\
  \bibinfo {note} {arXiv: 2007.11139}\BibitemShut {NoStop}%
\bibitem [{\citenamefont {Tran}\ and\ \citenamefont
  {Neuscamman}(2020)}]{tranImprovingExcitedState2020}%
  \BibitemOpen
  \bibfield  {author} {\bibinfo {author} {\bibfnamefont {L.~N.}\ \bibnamefont
  {Tran}}\ and\ \bibinfo {author} {\bibfnamefont {E.}~\bibnamefont
  {Neuscamman}},\ }\href {http://arxiv.org/abs/2006.09621} {\bibfield
  {journal} {\bibinfo  {journal} {arXiv:2006.09621 [physics]}\ } (\bibinfo
  {year} {2020})},\ \bibinfo {note} {arXiv: 2006.09621}\BibitemShut {NoStop}%
\bibitem [{\citenamefont {Ceperley}\ and\ \citenamefont
  {Bernu}(1988)}]{ceperleyCalculationExcitedState1988}%
  \BibitemOpen
  \bibfield  {author} {\bibinfo {author} {\bibfnamefont {D.~M.}\ \bibnamefont
  {Ceperley}}\ and\ \bibinfo {author} {\bibfnamefont {B.}~\bibnamefont
  {Bernu}},\ }\href {\doibase 10.1063/1.455398} {\bibfield  {journal} {\bibinfo
   {journal} {The Journal of Chemical Physics}\ }\textbf {\bibinfo {volume}
  {89}},\ \bibinfo {pages} {6316} (\bibinfo {year} {1988})},\ \bibinfo {note}
  {publisher: American Institute of Physics}\BibitemShut {NoStop}%
\bibitem [{\citenamefont {Blunt}\ \emph {et~al.}(2015)\citenamefont {Blunt},
  \citenamefont {Smart}, \citenamefont {Booth},\ and\ \citenamefont
  {Alavi}}]{bluntExcitedstateApproachFull2015}%
  \BibitemOpen
  \bibfield  {author} {\bibinfo {author} {\bibfnamefont {N.~S.}\ \bibnamefont
  {Blunt}}, \bibinfo {author} {\bibfnamefont {S.~D.}\ \bibnamefont {Smart}},
  \bibinfo {author} {\bibfnamefont {G.~H.}\ \bibnamefont {Booth}}, \ and\
  \bibinfo {author} {\bibfnamefont {A.}~\bibnamefont {Alavi}},\ }\href
  {\doibase 10.1063/1.4932595} {\bibfield  {journal} {\bibinfo  {journal} {The
  Journal of Chemical Physics}\ }\textbf {\bibinfo {volume} {143}},\ \bibinfo
  {pages} {134117} (\bibinfo {year} {2015})}\BibitemShut {NoStop}%
\bibitem [{\citenamefont {Filippi}\ \emph {et~al.}(2009)\citenamefont
  {Filippi}, \citenamefont {Zaccheddu},\ and\ \citenamefont
  {Buda}}]{doi:10.1021/ct900227j}%
  \BibitemOpen
  \bibfield  {author} {\bibinfo {author} {\bibfnamefont {C.}~\bibnamefont
  {Filippi}}, \bibinfo {author} {\bibfnamefont {M.}~\bibnamefont {Zaccheddu}},
  \ and\ \bibinfo {author} {\bibfnamefont {F.}~\bibnamefont {Buda}},\ }\href
  {\doibase 10.1021/ct900227j} {\bibfield  {journal} {\bibinfo  {journal}
  {Journal of Chemical Theory and Computation}\ }\textbf {\bibinfo {volume}
  {5}},\ \bibinfo {pages} {2074} (\bibinfo {year} {2009})},\ \bibinfo {note}
  {pMID: 26613149},\ \Eprint
  {http://arxiv.org/abs/https://doi.org/10.1021/ct900227j}
  {https://doi.org/10.1021/ct900227j} \BibitemShut {NoStop}%
\bibitem [{\citenamefont {Dash}\ \emph {et~al.}(2019)\citenamefont {Dash},
  \citenamefont {Feldt}, \citenamefont {Moroni}, \citenamefont {Scemama},\ and\
  \citenamefont {Filippi}}]{doi:10.1021/acs.jctc.9b00476}%
  \BibitemOpen
  \bibfield  {author} {\bibinfo {author} {\bibfnamefont {M.}~\bibnamefont
  {Dash}}, \bibinfo {author} {\bibfnamefont {J.}~\bibnamefont {Feldt}},
  \bibinfo {author} {\bibfnamefont {S.}~\bibnamefont {Moroni}}, \bibinfo
  {author} {\bibfnamefont {A.}~\bibnamefont {Scemama}}, \ and\ \bibinfo
  {author} {\bibfnamefont {C.}~\bibnamefont {Filippi}},\ }\href {\doibase
  10.1021/acs.jctc.9b00476} {\bibfield  {journal} {\bibinfo  {journal} {Journal
  of Chemical Theory and Computation}\ }\textbf {\bibinfo {volume} {15}},\
  \bibinfo {pages} {4896} (\bibinfo {year} {2019})},\ \bibinfo {note} {pMID:
  31348645},\ \Eprint
  {http://arxiv.org/abs/https://doi.org/10.1021/acs.jctc.9b00476}
  {https://doi.org/10.1021/acs.jctc.9b00476} \BibitemShut {NoStop}%
\bibitem [{\citenamefont {Send}\ \emph {et~al.}(2011)\citenamefont {Send},
  \citenamefont {Valsson},\ and\ \citenamefont
  {Filippi}}]{doi:10.1021/ct1006295}%
  \BibitemOpen
  \bibfield  {author} {\bibinfo {author} {\bibfnamefont {R.}~\bibnamefont
  {Send}}, \bibinfo {author} {\bibfnamefont {O.}~\bibnamefont {Valsson}}, \
  and\ \bibinfo {author} {\bibfnamefont {C.}~\bibnamefont {Filippi}},\ }\href
  {\doibase 10.1021/ct1006295} {\bibfield  {journal} {\bibinfo  {journal}
  {Journal of Chemical Theory and Computation}\ }\textbf {\bibinfo {volume}
  {7}},\ \bibinfo {pages} {444} (\bibinfo {year} {2011})},\ \bibinfo {note}
  {pMID: 26596164},\ \Eprint
  {http://arxiv.org/abs/https://doi.org/10.1021/ct1006295}
  {https://doi.org/10.1021/ct1006295} \BibitemShut {NoStop}%
\bibitem [{\citenamefont {Pineda~Flores}\ and\ \citenamefont
  {Neuscamman}(2019{\natexlab{a}})}]{pinedafloresExcitedStateSpecific2019}%
  \BibitemOpen
  \bibfield  {author} {\bibinfo {author} {\bibfnamefont {S.~D.}\ \bibnamefont
  {Pineda~Flores}}\ and\ \bibinfo {author} {\bibfnamefont {E.}~\bibnamefont
  {Neuscamman}},\ }\href {\doibase 10.1021/acs.jpca.8b10671} {\bibfield
  {journal} {\bibinfo  {journal} {The Journal of Physical Chemistry A}\
  }\textbf {\bibinfo {volume} {123}},\ \bibinfo {pages} {1487} (\bibinfo {year}
  {2019}{\natexlab{a}})}\BibitemShut {NoStop}%
\bibitem [{\citenamefont {Zhao}\ and\ \citenamefont
  {Neuscamman}(2016)}]{doi:10.1021/acs.jctc.6b00508}%
  \BibitemOpen
  \bibfield  {author} {\bibinfo {author} {\bibfnamefont {L.}~\bibnamefont
  {Zhao}}\ and\ \bibinfo {author} {\bibfnamefont {E.}~\bibnamefont
  {Neuscamman}},\ }\href {\doibase 10.1021/acs.jctc.6b00508} {\bibfield
  {journal} {\bibinfo  {journal} {Journal of Chemical Theory and Computation}\
  }\textbf {\bibinfo {volume} {12}},\ \bibinfo {pages} {3436} (\bibinfo {year}
  {2016})},\ \bibinfo {note} {pMID: 27379468},\ \Eprint
  {http://arxiv.org/abs/https://doi.org/10.1021/acs.jctc.6b00508}
  {https://doi.org/10.1021/acs.jctc.6b00508} \BibitemShut {NoStop}%
\bibitem [{\citenamefont {Shea}\ and\ \citenamefont
  {Neuscamman}(2017)}]{doi:10.1021/acs.jctc.7b00923}%
  \BibitemOpen
  \bibfield  {author} {\bibinfo {author} {\bibfnamefont {J.~A.~R.}\
  \bibnamefont {Shea}}\ and\ \bibinfo {author} {\bibfnamefont {E.}~\bibnamefont
  {Neuscamman}},\ }\href {\doibase 10.1021/acs.jctc.7b00923} {\bibfield
  {journal} {\bibinfo  {journal} {Journal of Chemical Theory and Computation}\
  }\textbf {\bibinfo {volume} {13}},\ \bibinfo {pages} {6078} (\bibinfo {year}
  {2017})},\ \bibinfo {note} {pMID: 29140699},\ \Eprint
  {http://arxiv.org/abs/https://doi.org/10.1021/acs.jctc.7b00923}
  {https://doi.org/10.1021/acs.jctc.7b00923} \BibitemShut {NoStop}%
\bibitem [{\citenamefont {Pineda~Flores}\ and\ \citenamefont
  {Neuscamman}(2019{\natexlab{b}})}]{doi:10.1021/acs.jpca.8b10671}%
  \BibitemOpen
  \bibfield  {author} {\bibinfo {author} {\bibfnamefont {S.~D.}\ \bibnamefont
  {Pineda~Flores}}\ and\ \bibinfo {author} {\bibfnamefont {E.}~\bibnamefont
  {Neuscamman}},\ }\href {\doibase 10.1021/acs.jpca.8b10671} {\bibfield
  {journal} {\bibinfo  {journal} {The Journal of Physical Chemistry A}\
  }\textbf {\bibinfo {volume} {123}},\ \bibinfo {pages} {1487} (\bibinfo {year}
  {2019}{\natexlab{b}})},\ \bibinfo {note} {pMID: 30702890},\ \Eprint
  {http://arxiv.org/abs/https://doi.org/10.1021/acs.jpca.8b10671}
  {https://doi.org/10.1021/acs.jpca.8b10671} \BibitemShut {NoStop}%
\bibitem [{\citenamefont {Blunt}\ and\ \citenamefont
  {Neuscamman}(2019)}]{doi:10.1021/acs.jctc.8b00879}%
  \BibitemOpen
  \bibfield  {author} {\bibinfo {author} {\bibfnamefont {N.~S.}\ \bibnamefont
  {Blunt}}\ and\ \bibinfo {author} {\bibfnamefont {E.}~\bibnamefont
  {Neuscamman}},\ }\href {\doibase 10.1021/acs.jctc.8b00879} {\bibfield
  {journal} {\bibinfo  {journal} {Journal of Chemical Theory and Computation}\
  }\textbf {\bibinfo {volume} {15}},\ \bibinfo {pages} {178} (\bibinfo {year}
  {2019})},\ \Eprint
  {http://arxiv.org/abs/https://doi.org/10.1021/acs.jctc.8b00879}
  {https://doi.org/10.1021/acs.jctc.8b00879} \BibitemShut {NoStop}%
\bibitem [{\citenamefont {Cuzzocrea}\ \emph {et~al.}(2020)\citenamefont
  {Cuzzocrea}, \citenamefont {Scemama}, \citenamefont {Briels}, \citenamefont
  {Moroni},\ and\ \citenamefont
  {Filippi}}]{cuzzocreaVariationalPrinciplesQuantum2020}%
  \BibitemOpen
  \bibfield  {author} {\bibinfo {author} {\bibfnamefont {A.}~\bibnamefont
  {Cuzzocrea}}, \bibinfo {author} {\bibfnamefont {A.}~\bibnamefont {Scemama}},
  \bibinfo {author} {\bibfnamefont {W.~J.}\ \bibnamefont {Briels}}, \bibinfo
  {author} {\bibfnamefont {S.}~\bibnamefont {Moroni}}, \ and\ \bibinfo {author}
  {\bibfnamefont {C.}~\bibnamefont {Filippi}},\ }\href {\doibase
  10.1021/acs.jctc.0c00147} {\bibfield  {journal} {\bibinfo  {journal} {Journal
  of Chemical Theory and Computation}\ }\textbf {\bibinfo {volume} {16}},\
  \bibinfo {pages} {4203} (\bibinfo {year} {2020})}\BibitemShut {NoStop}%
\bibitem [{\citenamefont {Choo}\ \emph {et~al.}(2018)\citenamefont {Choo},
  \citenamefont {Carleo}, \citenamefont {Regnault},\ and\ \citenamefont
  {Neupert}}]{chooSymmetriesManyBodyExcitations2018}%
  \BibitemOpen
  \bibfield  {author} {\bibinfo {author} {\bibfnamefont {K.}~\bibnamefont
  {Choo}}, \bibinfo {author} {\bibfnamefont {G.}~\bibnamefont {Carleo}},
  \bibinfo {author} {\bibfnamefont {N.}~\bibnamefont {Regnault}}, \ and\
  \bibinfo {author} {\bibfnamefont {T.}~\bibnamefont {Neupert}},\ }\href
  {\doibase 10.1103/PhysRevLett.121.167204} {\bibfield  {journal} {\bibinfo
  {journal} {Physical Review Letters}\ }\textbf {\bibinfo {volume} {121}},\
  \bibinfo {pages} {167204} (\bibinfo {year} {2018})}\BibitemShut {NoStop}%
\bibitem [{\citenamefont {Stoudenmire}\ and\ \citenamefont
  {White}(2012)}]{stoudenmireStudyingTwoDimensionalSystems2012}%
  \BibitemOpen
  \bibfield  {author} {\bibinfo {author} {\bibfnamefont {E.}~\bibnamefont
  {Stoudenmire}}\ and\ \bibinfo {author} {\bibfnamefont {S.~R.}\ \bibnamefont
  {White}},\ }\href {\doibase 10.1146/annurev-conmatphys-020911-125018}
  {\bibfield  {journal} {\bibinfo  {journal} {Annual Review of Condensed Matter
  Physics}\ }\textbf {\bibinfo {volume} {3}},\ \bibinfo {pages} {111} (\bibinfo
  {year} {2012})}\BibitemShut {NoStop}%
\bibitem [{pyq()}]{pyqmc}%
  \BibitemOpen
  \href@noop {} {\enquote {\bibinfo {title} {{PyQMC: Python library for real
  space quantum Monte Carlo}},}\ }\bibinfo {howpublished}
  {\url{https://github.com/WagnerGroup/pyqmc}}\BibitemShut {NoStop}%
\bibitem [{\citenamefont {Casula}\ \emph {et~al.}(2004)\citenamefont {Casula},
  \citenamefont {Attaccalite},\ and\ \citenamefont
  {Sorella}}]{casulaCorrelatedGeminalWave2004}%
  \BibitemOpen
  \bibfield  {author} {\bibinfo {author} {\bibfnamefont {M.}~\bibnamefont
  {Casula}}, \bibinfo {author} {\bibfnamefont {C.}~\bibnamefont {Attaccalite}},
  \ and\ \bibinfo {author} {\bibfnamefont {S.}~\bibnamefont {Sorella}},\ }\href
  {\doibase doi:10.1063/1.1794632} {\bibfield  {journal} {\bibinfo  {journal}
  {The Journal of Chemical Physics}\ }\textbf {\bibinfo {volume} {121}},\
  \bibinfo {pages} {7110} (\bibinfo {year} {2004})}\BibitemShut {NoStop}%
\bibitem [{Note1()}]{Note1}%
  \BibitemOpen
  \bibinfo {note} {Since we are using multi-Slater-Jastrow wave functions in
  this work, it was more convenient to control the normalization of the wave
  function in this way. Another option, not tried here, is to parameterize the
  wave function in a norm-conserving manner, which could increase the
  efficiency of the algorithm. On the other hand, for emerging wave functions,
  it may be more convenient to allow the normalization to vary.}\BibitemShut
  {Stop}%
\bibitem [{\citenamefont {Wagner}\ \emph {et~al.}(2009)\citenamefont {Wagner},
  \citenamefont {Bajdich},\ and\ \citenamefont {Mitas}}]{WAGNER20093390}%
  \BibitemOpen
  \bibfield  {author} {\bibinfo {author} {\bibfnamefont {L.~K.}\ \bibnamefont
  {Wagner}}, \bibinfo {author} {\bibfnamefont {M.}~\bibnamefont {Bajdich}}, \
  and\ \bibinfo {author} {\bibfnamefont {L.}~\bibnamefont {Mitas}},\ }\href
  {\doibase https://doi.org/10.1016/j.jcp.2009.01.017} {\bibfield  {journal}
  {\bibinfo  {journal} {Journal of Computational Physics}\ }\textbf {\bibinfo
  {volume} {228}},\ \bibinfo {pages} {3390 } (\bibinfo {year}
  {2009})}\BibitemShut {NoStop}%
\bibitem [{\citenamefont {Koster}\ and\ \citenamefont
  {Rahmann}(2012)}]{kosterSnakemakeScalableBioinformatics2012}%
  \BibitemOpen
  \bibfield  {author} {\bibinfo {author} {\bibfnamefont {J.}~\bibnamefont
  {Koster}}\ and\ \bibinfo {author} {\bibfnamefont {S.}~\bibnamefont
  {Rahmann}},\ }\href {\doibase 10.1093/bioinformatics/bts480} {\bibfield
  {journal} {\bibinfo  {journal} {Bioinformatics}\ }\textbf {\bibinfo {volume}
  {28}},\ \bibinfo {pages} {2520} (\bibinfo {year} {2012})}\BibitemShut
  {NoStop}%
\bibitem [{\citenamefont {Sun}\ \emph {et~al.}(2018)\citenamefont {Sun},
  \citenamefont {Berkelbach}, \citenamefont {Blunt}, \citenamefont {Booth},
  \citenamefont {Guo}, \citenamefont {Li}, \citenamefont {Liu}, \citenamefont
  {McClain}, \citenamefont {Sayfutyarova}, \citenamefont {Sharma},
  \citenamefont {Wouters},\ and\ \citenamefont {Chan}}]{doi:10.1002/wcms.1340}%
  \BibitemOpen
  \bibfield  {author} {\bibinfo {author} {\bibfnamefont {Q.}~\bibnamefont
  {Sun}}, \bibinfo {author} {\bibfnamefont {T.~C.}\ \bibnamefont {Berkelbach}},
  \bibinfo {author} {\bibfnamefont {N.~S.}\ \bibnamefont {Blunt}}, \bibinfo
  {author} {\bibfnamefont {G.~H.}\ \bibnamefont {Booth}}, \bibinfo {author}
  {\bibfnamefont {S.}~\bibnamefont {Guo}}, \bibinfo {author} {\bibfnamefont
  {Z.}~\bibnamefont {Li}}, \bibinfo {author} {\bibfnamefont {J.}~\bibnamefont
  {Liu}}, \bibinfo {author} {\bibfnamefont {J.~D.}\ \bibnamefont {McClain}},
  \bibinfo {author} {\bibfnamefont {E.~R.}\ \bibnamefont {Sayfutyarova}},
  \bibinfo {author} {\bibfnamefont {S.}~\bibnamefont {Sharma}}, \bibinfo
  {author} {\bibfnamefont {S.}~\bibnamefont {Wouters}}, \ and\ \bibinfo
  {author} {\bibfnamefont {G.~K.-L.}\ \bibnamefont {Chan}},\ }\href {\doibase
  10.1002/wcms.1340} {\bibfield  {journal} {\bibinfo  {journal} {WIREs
  Computational Molecular Science}\ }\textbf {\bibinfo {volume} {8}},\ \bibinfo
  {pages} {e1340} (\bibinfo {year} {2018})},\ \Eprint
  {http://arxiv.org/abs/https://onlinelibrary.wiley.com/doi/pdf/10.1002/wcms.1340}
  {https://onlinelibrary.wiley.com/doi/pdf/10.1002/wcms.1340} \BibitemShut
  {NoStop}%
\bibitem [{\citenamefont {Dunning}(1989)}]{dunningGaussianBasisSets1989}%
  \BibitemOpen
  \bibfield  {author} {\bibinfo {author} {\bibfnamefont {T.~H.}\ \bibnamefont
  {Dunning}},\ }\href {\doibase 10.1063/1.456153} {\bibfield  {journal}
  {\bibinfo  {journal} {The Journal of Chemical Physics}\ }\textbf {\bibinfo
  {volume} {90}},\ \bibinfo {pages} {1007} (\bibinfo {year}
  {1989})}\BibitemShut {NoStop}%
\bibitem [{\citenamefont {Roos}\ \emph {et~al.}(1992)\citenamefont {Roos},
  \citenamefont {Andersson},\ and\ \citenamefont {Fülscher}}]{ROOS19925}%
  \BibitemOpen
  \bibfield  {author} {\bibinfo {author} {\bibfnamefont {B.~O.}\ \bibnamefont
  {Roos}}, \bibinfo {author} {\bibfnamefont {K.}~\bibnamefont {Andersson}}, \
  and\ \bibinfo {author} {\bibfnamefont {M.~P.}\ \bibnamefont {Fülscher}},\
  }\href {\doibase https://doi.org/10.1016/0009-2614(92)85419-B} {\bibfield
  {journal} {\bibinfo  {journal} {Chemical Physics Letters}\ }\textbf {\bibinfo
  {volume} {192}},\ \bibinfo {pages} {5 } (\bibinfo {year} {1992})}\BibitemShut
  {NoStop}%
\bibitem [{\citenamefont {Christiansen}\ \emph {et~al.}(1996)\citenamefont
  {Christiansen}, \citenamefont {Koch}, \citenamefont {Halkier}, \citenamefont
  {Jo/rgensen}, \citenamefont {Helgaker},\ and\ \citenamefont {Sánchez~de
  Merás}}]{doi:10.1063/1.471985}%
  \BibitemOpen
  \bibfield  {author} {\bibinfo {author} {\bibfnamefont {O.}~\bibnamefont
  {Christiansen}}, \bibinfo {author} {\bibfnamefont {H.}~\bibnamefont {Koch}},
  \bibinfo {author} {\bibfnamefont {A.}~\bibnamefont {Halkier}}, \bibinfo
  {author} {\bibfnamefont {P.}~\bibnamefont {Jo/rgensen}}, \bibinfo {author}
  {\bibfnamefont {T.}~\bibnamefont {Helgaker}}, \ and\ \bibinfo {author}
  {\bibfnamefont {A.}~\bibnamefont {Sánchez~de Merás}},\ }\href {\doibase
  10.1063/1.471985} {\bibfield  {journal} {\bibinfo  {journal} {The Journal of
  Chemical Physics}\ }\textbf {\bibinfo {volume} {105}},\ \bibinfo {pages}
  {6921} (\bibinfo {year} {1996})},\ \Eprint
  {http://arxiv.org/abs/https://doi.org/10.1063/1.471985}
  {https://doi.org/10.1063/1.471985} \BibitemShut {NoStop}%
\bibitem [{\citenamefont {Schreiber}\ \emph {et~al.}(2008)\citenamefont
  {Schreiber}, \citenamefont {Silva-Junior}, \citenamefont {Sauer},\ and\
  \citenamefont {Thiel}}]{doi:10.1063/1.2889385}%
  \BibitemOpen
  \bibfield  {author} {\bibinfo {author} {\bibfnamefont {M.}~\bibnamefont
  {Schreiber}}, \bibinfo {author} {\bibfnamefont {M.~R.}\ \bibnamefont
  {Silva-Junior}}, \bibinfo {author} {\bibfnamefont {S.~P.~A.}\ \bibnamefont
  {Sauer}}, \ and\ \bibinfo {author} {\bibfnamefont {W.}~\bibnamefont
  {Thiel}},\ }\href {\doibase 10.1063/1.2889385} {\bibfield  {journal}
  {\bibinfo  {journal} {The Journal of Chemical Physics}\ }\textbf {\bibinfo
  {volume} {128}},\ \bibinfo {pages} {134110} (\bibinfo {year} {2008})},\
  \Eprint {http://arxiv.org/abs/https://doi.org/10.1063/1.2889385}
  {https://doi.org/10.1063/1.2889385} \BibitemShut {NoStop}%
\bibitem [{\citenamefont {Burkatzki}\ \emph {et~al.}(2007)\citenamefont
  {Burkatzki}, \citenamefont {Filippi},\ and\ \citenamefont
  {Dolg}}]{doi:10.1063/1.2741534}%
  \BibitemOpen
  \bibfield  {author} {\bibinfo {author} {\bibfnamefont {M.}~\bibnamefont
  {Burkatzki}}, \bibinfo {author} {\bibfnamefont {C.}~\bibnamefont {Filippi}},
  \ and\ \bibinfo {author} {\bibfnamefont {M.}~\bibnamefont {Dolg}},\ }\href
  {\doibase 10.1063/1.2741534} {\bibfield  {journal} {\bibinfo  {journal} {The
  Journal of Chemical Physics}\ }\textbf {\bibinfo {volume} {126}},\ \bibinfo
  {pages} {234105} (\bibinfo {year} {2007})},\ \Eprint
  {http://arxiv.org/abs/https://doi.org/10.1063/1.2741534}
  {https://doi.org/10.1063/1.2741534} \BibitemShut {NoStop}%
\bibitem [{\citenamefont {Shull}(1949)}]{doi:10.1063/1.1747242}%
  \BibitemOpen
  \bibfield  {author} {\bibinfo {author} {\bibfnamefont {H.}~\bibnamefont
  {Shull}},\ }\href {\doibase 10.1063/1.1747242} {\bibfield  {journal}
  {\bibinfo  {journal} {The Journal of Chemical Physics}\ }\textbf {\bibinfo
  {volume} {17}},\ \bibinfo {pages} {295} (\bibinfo {year} {1949})},\ \Eprint
  {http://arxiv.org/abs/https://doi.org/10.1063/1.1747242}
  {https://doi.org/10.1063/1.1747242} \BibitemShut {NoStop}%
\bibitem [{\citenamefont {Doering}(1977)}]{doi:10.1063/1.435381}%
  \BibitemOpen
  \bibfield  {author} {\bibinfo {author} {\bibfnamefont {J.~P.}\ \bibnamefont
  {Doering}},\ }\href {\doibase 10.1063/1.435381} {\bibfield  {journal}
  {\bibinfo  {journal} {The Journal of Chemical Physics}\ }\textbf {\bibinfo
  {volume} {67}},\ \bibinfo {pages} {4065} (\bibinfo {year} {1977})},\ \Eprint
  {http://arxiv.org/abs/https://aip.scitation.org/doi/pdf/10.1063/1.435381}
  {https://aip.scitation.org/doi/pdf/10.1063/1.435381} \BibitemShut {NoStop}%
\bibitem [{\citenamefont {Christiansen}\ \emph {et~al.}(1998)\citenamefont
  {Christiansen}, \citenamefont {Stanton},\ and\ \citenamefont
  {Gauss}}]{doi:10.1063/1.475801}%
  \BibitemOpen
  \bibfield  {author} {\bibinfo {author} {\bibfnamefont {O.}~\bibnamefont
  {Christiansen}}, \bibinfo {author} {\bibfnamefont {J.~F.}\ \bibnamefont
  {Stanton}}, \ and\ \bibinfo {author} {\bibfnamefont {J.}~\bibnamefont
  {Gauss}},\ }\href {\doibase 10.1063/1.475801} {\bibfield  {journal} {\bibinfo
   {journal} {The Journal of Chemical Physics}\ }\textbf {\bibinfo {volume}
  {108}},\ \bibinfo {pages} {3987} (\bibinfo {year} {1998})},\ \Eprint
  {http://arxiv.org/abs/https://doi.org/10.1063/1.475801}
  {https://doi.org/10.1063/1.475801} \BibitemShut {NoStop}%
\bibitem [{\citenamefont {Bursill}\ \emph {et~al.}(1998)\citenamefont
  {Bursill}, \citenamefont {Castleton},\ and\ \citenamefont
  {Barford}}]{BURSILL1998305}%
  \BibitemOpen
  \bibfield  {author} {\bibinfo {author} {\bibfnamefont {R.~J.}\ \bibnamefont
  {Bursill}}, \bibinfo {author} {\bibfnamefont {C.}~\bibnamefont {Castleton}},
  \ and\ \bibinfo {author} {\bibfnamefont {W.}~\bibnamefont {Barford}},\ }\href
  {\doibase https://doi.org/10.1016/S0009-2614(98)00903-8} {\bibfield
  {journal} {\bibinfo  {journal} {Chemical Physics Letters}\ }\textbf {\bibinfo
  {volume} {294}},\ \bibinfo {pages} {305 } (\bibinfo {year}
  {1998})}\BibitemShut {NoStop}%
\bibitem [{\citenamefont {Adamo}\ \emph {et~al.}(1999)\citenamefont {Adamo},
  \citenamefont {Scuseria},\ and\ \citenamefont
  {Barone}}]{doi:10.1063/1.479571}%
  \BibitemOpen
  \bibfield  {author} {\bibinfo {author} {\bibfnamefont {C.}~\bibnamefont
  {Adamo}}, \bibinfo {author} {\bibfnamefont {G.~E.}\ \bibnamefont {Scuseria}},
  \ and\ \bibinfo {author} {\bibfnamefont {V.}~\bibnamefont {Barone}},\ }\href
  {\doibase 10.1063/1.479571} {\bibfield  {journal} {\bibinfo  {journal} {The
  Journal of Chemical Physics}\ }\textbf {\bibinfo {volume} {111}},\ \bibinfo
  {pages} {2889} (\bibinfo {year} {1999})},\ \Eprint
  {http://arxiv.org/abs/https://doi.org/10.1063/1.479571}
  {https://doi.org/10.1063/1.479571} \BibitemShut {NoStop}%
\bibitem [{\citenamefont {Ziegler}\ and\ \citenamefont
  {Hudson}(1982)}]{Ziegler1982TheVS}%
  \BibitemOpen
  \bibfield  {author} {\bibinfo {author} {\bibfnamefont {L.}~\bibnamefont
  {Ziegler}}\ and\ \bibinfo {author} {\bibfnamefont {B.}~\bibnamefont
  {Hudson}}\ }(\bibinfo {year} {1982})\BibitemShut {NoStop}%
\bibitem [{\citenamefont {Hajgató}\ \emph {et~al.}(2009)\citenamefont
  {Hajgató}, \citenamefont {Szieberth}, \citenamefont {Geerlings},
  \citenamefont {De~Proft},\ and\ \citenamefont
  {Deleuze}}]{doi:10.1063/1.3270190}%
  \BibitemOpen
  \bibfield  {author} {\bibinfo {author} {\bibfnamefont {B.}~\bibnamefont
  {Hajgató}}, \bibinfo {author} {\bibfnamefont {D.}~\bibnamefont {Szieberth}},
  \bibinfo {author} {\bibfnamefont {P.}~\bibnamefont {Geerlings}}, \bibinfo
  {author} {\bibfnamefont {F.}~\bibnamefont {De~Proft}}, \ and\ \bibinfo
  {author} {\bibfnamefont {M.~S.}\ \bibnamefont {Deleuze}},\ }\href {\doibase
  10.1063/1.3270190} {\bibfield  {journal} {\bibinfo  {journal} {The Journal of
  Chemical Physics}\ }\textbf {\bibinfo {volume} {131}},\ \bibinfo {pages}
  {224321} (\bibinfo {year} {2009})},\ \Eprint
  {http://arxiv.org/abs/https://doi.org/10.1063/1.3270190}
  {https://doi.org/10.1063/1.3270190} \BibitemShut {NoStop}%
\bibitem [{\citenamefont {Ohno}\ and\ \citenamefont
  {Takahashi}(2002)}]{OHNO2002409}%
  \BibitemOpen
  \bibfield  {author} {\bibinfo {author} {\bibfnamefont {K.}~\bibnamefont
  {Ohno}}\ and\ \bibinfo {author} {\bibfnamefont {R.}~\bibnamefont
  {Takahashi}},\ }\href {\doibase
  https://doi.org/10.1016/S0009-2614(02)00399-8} {\bibfield  {journal}
  {\bibinfo  {journal} {Chemical Physics Letters}\ }\textbf {\bibinfo {volume}
  {356}},\ \bibinfo {pages} {409 } (\bibinfo {year} {2002})}\BibitemShut
  {NoStop}%
\bibitem [{\citenamefont {Loos}\ \emph {et~al.}(2019)\citenamefont {Loos},
  \citenamefont {Boggio-Pasqua}, \citenamefont {Scemama}, \citenamefont
  {Caffarel},\ and\ \citenamefont
  {Jacquemin}}]{loosReferenceEnergiesDouble2019}%
  \BibitemOpen
  \bibfield  {author} {\bibinfo {author} {\bibfnamefont {P.-F.}\ \bibnamefont
  {Loos}}, \bibinfo {author} {\bibfnamefont {M.}~\bibnamefont {Boggio-Pasqua}},
  \bibinfo {author} {\bibfnamefont {A.}~\bibnamefont {Scemama}}, \bibinfo
  {author} {\bibfnamefont {M.}~\bibnamefont {Caffarel}}, \ and\ \bibinfo
  {author} {\bibfnamefont {D.}~\bibnamefont {Jacquemin}},\ }\href {\doibase
  10.1021/acs.jctc.8b01205} {\bibfield  {journal} {\bibinfo  {journal} {Journal
  of Chemical Theory and Computation}\ }\textbf {\bibinfo {volume} {15}},\
  \bibinfo {pages} {1939} (\bibinfo {year} {2019})}\BibitemShut {NoStop}%
\bibitem [{\citenamefont {Kwon}\ \emph {et~al.}(1993)\citenamefont {Kwon},
  \citenamefont {Ceperley},\ and\ \citenamefont
  {Martin}}]{kwonEffectsThreebodyBackflow1993}%
  \BibitemOpen
  \bibfield  {author} {\bibinfo {author} {\bibfnamefont {Y.}~\bibnamefont
  {Kwon}}, \bibinfo {author} {\bibfnamefont {D.~M.}\ \bibnamefont {Ceperley}},
  \ and\ \bibinfo {author} {\bibfnamefont {R.~M.}\ \bibnamefont {Martin}},\
  }\href {\doibase 10.1103/PhysRevB.48.12037} {\bibfield  {journal} {\bibinfo
  {journal} {Physical Review B}\ }\textbf {\bibinfo {volume} {48}},\ \bibinfo
  {pages} {12037} (\bibinfo {year} {1993})}\BibitemShut {NoStop}%
\bibitem [{\citenamefont {López~Ríos}\ \emph {et~al.}(2006)\citenamefont
  {López~Ríos}, \citenamefont {Ma}, \citenamefont {Drummond}, \citenamefont
  {Towler},\ and\ \citenamefont
  {Needs}}]{lopezriosInhomogeneousBackflowTransformations2006a}%
  \BibitemOpen
  \bibfield  {author} {\bibinfo {author} {\bibfnamefont {P.}~\bibnamefont
  {López~Ríos}}, \bibinfo {author} {\bibfnamefont {A.}~\bibnamefont {Ma}},
  \bibinfo {author} {\bibfnamefont {N.~D.}\ \bibnamefont {Drummond}}, \bibinfo
  {author} {\bibfnamefont {M.~D.}\ \bibnamefont {Towler}}, \ and\ \bibinfo
  {author} {\bibfnamefont {R.~J.}\ \bibnamefont {Needs}},\ }\href {\doibase
  10.1103/PhysRevE.74.066701} {\bibfield  {journal} {\bibinfo  {journal}
  {Physical Review E}\ }\textbf {\bibinfo {volume} {74}},\ \bibinfo {pages}
  {066701} (\bibinfo {year} {2006})}\BibitemShut {NoStop}%
\bibitem [{\citenamefont {Tocchio}\ \emph {et~al.}(2008)\citenamefont
  {Tocchio}, \citenamefont {Becca}, \citenamefont {Parola},\ and\ \citenamefont
  {Sorella}}]{PhysRevB.78.041101}%
  \BibitemOpen
  \bibfield  {author} {\bibinfo {author} {\bibfnamefont {L.~F.}\ \bibnamefont
  {Tocchio}}, \bibinfo {author} {\bibfnamefont {F.}~\bibnamefont {Becca}},
  \bibinfo {author} {\bibfnamefont {A.}~\bibnamefont {Parola}}, \ and\ \bibinfo
  {author} {\bibfnamefont {S.}~\bibnamefont {Sorella}},\ }\href {\doibase
  10.1103/PhysRevB.78.041101} {\bibfield  {journal} {\bibinfo  {journal} {Phys.
  Rev. B}\ }\textbf {\bibinfo {volume} {78}},\ \bibinfo {pages} {041101}
  (\bibinfo {year} {2008})}\BibitemShut {NoStop}%
\bibitem [{\citenamefont {Luo}\ and\ \citenamefont
  {Clark}(2019)}]{PhysRevLett.122.226401}%
  \BibitemOpen
  \bibfield  {author} {\bibinfo {author} {\bibfnamefont {D.}~\bibnamefont
  {Luo}}\ and\ \bibinfo {author} {\bibfnamefont {B.~K.}\ \bibnamefont
  {Clark}},\ }\href {\doibase 10.1103/PhysRevLett.122.226401} {\bibfield
  {journal} {\bibinfo  {journal} {Phys. Rev. Lett.}\ }\textbf {\bibinfo
  {volume} {122}},\ \bibinfo {pages} {226401} (\bibinfo {year}
  {2019})}\BibitemShut {NoStop}%
\bibitem [{\citenamefont {Pfau}\ \emph {et~al.}(2020)\citenamefont {Pfau},
  \citenamefont {Spencer}, \citenamefont {Matthews},\ and\ \citenamefont
  {Foulkes}}]{pfauInitioSolutionManyelectron2020}%
  \BibitemOpen
  \bibfield  {author} {\bibinfo {author} {\bibfnamefont {D.}~\bibnamefont
  {Pfau}}, \bibinfo {author} {\bibfnamefont {J.~S.}\ \bibnamefont {Spencer}},
  \bibinfo {author} {\bibfnamefont {A.~G. D.~G.}\ \bibnamefont {Matthews}}, \
  and\ \bibinfo {author} {\bibfnamefont {W.~M.~C.}\ \bibnamefont {Foulkes}},\
  }\href {\doibase 10.1103/PhysRevResearch.2.033429} {\bibfield  {journal}
  {\bibinfo  {journal} {Physical Review Research}\ }\textbf {\bibinfo {volume}
  {2}},\ \bibinfo {pages} {033429} (\bibinfo {year} {2020})}\BibitemShut
  {NoStop}%
\bibitem [{\citenamefont {Carleo}\ and\ \citenamefont
  {Troyer}(2017)}]{carleoSolvingQuantumManybody2017}%
  \BibitemOpen
  \bibfield  {author} {\bibinfo {author} {\bibfnamefont {G.}~\bibnamefont
  {Carleo}}\ and\ \bibinfo {author} {\bibfnamefont {M.}~\bibnamefont
  {Troyer}},\ }\href {\doibase 10.1126/science.aag2302} {\bibfield  {journal}
  {\bibinfo  {journal} {Science}\ }\textbf {\bibinfo {volume} {355}},\ \bibinfo
  {pages} {602} (\bibinfo {year} {2017})}\BibitemShut {NoStop}%
\bibitem [{\citenamefont {Bajdich}\ \emph {et~al.}(2008)\citenamefont
  {Bajdich}, \citenamefont {Mitas}, \citenamefont {Wagner},\ and\ \citenamefont
  {Schmidt}}]{PhysRevB.77.115112}%
  \BibitemOpen
  \bibfield  {author} {\bibinfo {author} {\bibfnamefont {M.}~\bibnamefont
  {Bajdich}}, \bibinfo {author} {\bibfnamefont {L.}~\bibnamefont {Mitas}},
  \bibinfo {author} {\bibfnamefont {L.~K.}\ \bibnamefont {Wagner}}, \ and\
  \bibinfo {author} {\bibfnamefont {K.~E.}\ \bibnamefont {Schmidt}},\ }\href
  {\doibase 10.1103/PhysRevB.77.115112} {\bibfield  {journal} {\bibinfo
  {journal} {Phys. Rev. B}\ }\textbf {\bibinfo {volume} {77}},\ \bibinfo
  {pages} {115112} (\bibinfo {year} {2008})}\BibitemShut {NoStop}%
\bibitem [{\citenamefont {Zheng}\ and\ \citenamefont
  {Wagner}(2015{\natexlab{b}})}]{PhysRevLett.114.176401}%
  \BibitemOpen
  \bibfield  {author} {\bibinfo {author} {\bibfnamefont {H.}~\bibnamefont
  {Zheng}}\ and\ \bibinfo {author} {\bibfnamefont {L.~K.}\ \bibnamefont
  {Wagner}},\ }\href {\doibase 10.1103/PhysRevLett.114.176401} {\bibfield
  {journal} {\bibinfo  {journal} {Phys. Rev. Lett.}\ }\textbf {\bibinfo
  {volume} {114}},\ \bibinfo {pages} {176401} (\bibinfo {year}
  {2015}{\natexlab{b}})}\BibitemShut {NoStop}%
\bibitem [{\citenamefont {Zheng}\ \emph {et~al.}(2018)\citenamefont {Zheng},
  \citenamefont {Changlani}, \citenamefont {Williams}, \citenamefont
  {Busemeyer},\ and\ \citenamefont {Wagner}}]{10.3389/fphy.2018.00043}%
  \BibitemOpen
  \bibfield  {author} {\bibinfo {author} {\bibfnamefont {H.}~\bibnamefont
  {Zheng}}, \bibinfo {author} {\bibfnamefont {H.~J.}\ \bibnamefont
  {Changlani}}, \bibinfo {author} {\bibfnamefont {K.~T.}\ \bibnamefont
  {Williams}}, \bibinfo {author} {\bibfnamefont {B.}~\bibnamefont {Busemeyer}},
  \ and\ \bibinfo {author} {\bibfnamefont {L.~K.}\ \bibnamefont {Wagner}},\
  }\href {\doibase 10.3389/fphy.2018.00043} {\bibfield  {journal} {\bibinfo
  {journal} {Frontiers in Physics}\ }\textbf {\bibinfo {volume} {6}},\ \bibinfo
  {pages} {43} (\bibinfo {year} {2018})}\BibitemShut {NoStop}%
\bibitem [{\citenamefont
  {MacDonald}(1933)}]{macdonaldSuccessiveApproximationsRayleighRitz1933}%
  \BibitemOpen
  \bibfield  {author} {\bibinfo {author} {\bibfnamefont {J.~K.~L.}\
  \bibnamefont {MacDonald}},\ }\href {\doibase 10.1103/PhysRev.43.830}
  {\bibfield  {journal} {\bibinfo  {journal} {Physical Review}\ }\textbf
  {\bibinfo {volume} {43}},\ \bibinfo {pages} {830} (\bibinfo {year} {1933})},\
  \bibinfo {note} {publisher: American Physical Society}\BibitemShut {NoStop}%
\end{thebibliography}%

\appendix

\section{Properties of $E(|\braket{\Psi|\Phi_0}|^2)$, and the approximate version$E(|\braket{\Psi|\tilde{\Phi}_0}|^2)$ }

Here we present a few simple properties of the energy functional for reference.

\begin{enumerate}
    \item $E(|\braket{\Psi|\Phi_i}|^2) \ge E_0 - (E_1-E_0) |\braket{\Psi|\Phi_i}|^2$. This is the bounding line.
    \item For any complete linear subspace, the minimization will form a line, and there are variational upper bounds.\cite{macdonaldSuccessiveApproximationsRayleighRitz1933}
    \item Broken symmetry wave function parameterizations (such as unrestricted Slater-Jastrow wave functions) do not usually comprise a complete linear space, so they may not form a line. 
\end{enumerate}

\end{document}